\documentclass[11pt,a4paper]{article}
\usepackage{cite}

\usepackage{amsmath, amsthm,mathtools, url,amssymb,upgreek,slashed}
\usepackage{ifpdf}
\ifpdf
  \usepackage[pdftex]{graphicx}
  \usepackage{epstopdf}
\else
  \usepackage[dvips]{graphicx}
\fi
\parskip=\baselineskip

\usepackage{comment}

\usepackage{hyperref}

\usepackage{xcolor}

\newcommand\blue[1]{{#1}}

\begin{document}
\begin{titlepage}
\begin{flushright}

\end{flushright}

\vskip 1.5in
\begin{center}
{\bf\Large{Averaging Over Narain Moduli Space}}
\vskip
0.5cm { Alexander Maloney} \vskip 0.05in {\small{ \textit{Department of Physics, McGill University}\vskip -.4cm
{\textit{ Montreal, QC H3A 2T8, Canada}}}
}
\vskip
0.5cm { Edward Witten} \vskip 0.05in {\small{ \textit{Institute for Advanced Study}\vskip -.4cm
{\textit{Einstein Drive, Princeton, NJ 08540 USA}}}
}
\end{center}
\vskip 0.5in
\baselineskip 16pt
\begin{abstract}   Recent developments involving JT gravity in two dimensions indicate that under some conditions, a gravitational path integral
is dual to an average over an ensemble of boundary theories, rather than to a specific boundary theory.   For an example in one dimension more,
one would like to compare a random ensemble of two-dimensional CFT's to Einstein gravity in three dimensions.  But this is difficult.   For a simpler
problem, here we average over Narain's family of two-dimensional CFT's obtained by toroidal compactification.
These theories are believed to be the most general ones with their central charges and abelian current algebra symmetries, so averaging over them means picking
a random CFT with those properties. The average can be computed using the Siegel-Weil formula of number theory and 
 has some properties suggestive of a bulk dual theory that
would be an exotic theory of gravity in three dimensions.    The bulk dual theory would be more like ${\mathrm U}(1)^{2D}$ Chern-Simons theory than like Einstein
gravity.  \end{abstract}
\date{May, 2020}
\end{titlepage}
\def\SO{{\mathrm{SO}}}
\def\G{{\text{\sf G}}}
\def\la{\langle}
\def\ra{\rangle}
\def\g{{\cmmib g}}
\def\U{{\mathrm U}}
\def\M{{\mathcal M}}
\def\d{{\mathrm d}}
\def\CS{{\mathrm{CS}}}
\def\Z{{\Bbb Z}}
\def\R{{\Bbb R}}
\def\J{{\mathcal J}}
\def\Bbb{\mathbb}
\def\Tr{{\rm Tr}}
\def\16{{\bf 16}}
\def\1{{(1)}}
\def\2{{(2)}}
\def\3{{\bf 3}}
\def\4{{\bf 4}}
\def\sg{{\mathrm g}}
\def\i{{\mathrm i}}
\def\h{\widehat}
\def\u{u}
\def\D{D}
\def\sp{{\sigma}}
\def\E{{\mathcal E}}
\def\O{{\mathcal O}}
\def\PF{{\mathit{P}\negthinspace\mathit{F}}}
\def\tr{{\mathrm{tr}}}
\def\be{\begin{equation}}
\def\ee{\end{equation}}
 \def\Sp{{\mathrm{Sp}}}
 \def\Spin{{\mathrm{Spin}}}
 \def\SL{{\mathrm{SL}}}
 \def\SU{{\mathrm{SU}}}
 \def\SO{{\mathrm{SO}}}
 \def\ll{\langle\langle}
\def\rr{\rangle\rangle}
\def\la{\langle}
\def\ra{\rangle}
\def\T{{\mathcal T}}
\def\V{{\mathcal V}}
\def\bar{\overline}
\def\v{v}
\def\dD{{\sf D}}
\def\tilde{\widetilde}
\def\t{\widetilde}
\def\R{{\Bbb{R}}}\def\Z{{\Bbb{Z}}}
\def\N{{\mathcal N}}
\def\B{{\mathcal B}}
\def\H{{\mathcal H}}
\def\hat{\widehat}
\def\Pf{{\mathrm{Pf}}}
\def\PSL{{\mathrm{PSL}}}
\def\Im{{\mathrm{Im}}}
\font\teneurm=eurm10 \font\seveneurm=eurm7 \font\fiveeurm=eurm5
\newfam\eurmfam
\textfont\eurmfam=\teneurm \scriptfont\eurmfam=\seveneurm
\scriptscriptfont\eurmfam=\fiveeurm
\def\eurm#1{{\fam\eurmfam\relax#1}}
\font\teneusm=eusm10 \font\seveneusm=eusm7 \font\fiveeusm=eusm5
\newfam\eusmfam
\textfont\eusmfam=\teneusm \scriptfont\eusmfam=\seveneusm
\scriptscriptfont\eusmfam=\fiveeusm
\def\eusm#1{{\fam\eusmfam\relax#1}}
\font\tencmmib=cmmib10 \skewchar\tencmmib='177
\font\sevencmmib=cmmib7 \skewchar\sevencmmib='177
\font\fivecmmib=cmmib5 \skewchar\fivecmmib='177
\newfam\cmmibfam
\textfont\cmmibfam=\tencmmib \scriptfont\cmmibfam=\sevencmmib
\scriptscriptfont\cmmibfam=\fivecmmib
\def\cmmib#1{{\fam\cmmibfam\relax#1}}
\numberwithin{equation}{section}
\def\a{{\eusm A}}
\def\b{{\eusm B}}
\def\neg{\negthinspace}
\def\d{\mathrm d}
\def\C{{\Bbb C}}
\def\HH{{\mathbb H}}
\def\P{{\mathcal P}}
\def\NS{{\sf{NS}}}
\def\Ra{{\sf{R}}}
\def\sV{{\sf V}}
\def\Z{{\Bbb Z}}
\def\A{{\eusm A}}
\def\B{{\eusm B}}
\def\S{{\mathcal S}}
\def\bar{\overline}
\def\sc{{\mathrm{sc}}}
\def\Max{{\mathrm{Max}}}
\def\CS{{\mathrm{CS}}}
\def\ga{\gamma}
\def\bg{\bar\ga}
\def\W{{\mathcal W}}
\def\M{{\mathcal M}}
\def\bM{{\overline \M}}
\def\L{{\mathcal L}}
\def\sM{{\sf M}}
\def\gst{\mathrm{g}_{\mathrm{st}}}
\def\gstt{\widetilde{\mathrm{g}}_{\mathrm{st}}}
\def\W{{W}}

\def\be{\begin{equation}}
\def\ee{\end{equation}}

\tableofcontents

\section{Introduction}\label{intro}
 
A simple model of gravity in two dimensions -- JT gravity -- is  dual to a random ensemble
 of quantum mechanical systems, rather than a specific quantum mechanical system \cite{SSS}.  It is natural to wonder if something similar
 happens in higher dimensions.    For example, gravity is still relatively simple in three spacetime dimensions, at least from some points of
 view.   Are there simple theories of gravity in three dimensions -- maybe even pure Einstein gravity -- that are dual in some sense to
 a random two-dimensional conformal field theory (CFT)?
 
 The difficulty here is that while a quantum mechanical system can be defined by specifying a Hamiltonian, the data required to specify
 a 2d CFT are far more complicated.    Accordingly, it is far from clear what should be meant by a random 2d CFT, though one can
 possibly get some insight from results about asymptotic behavior of dimensions and couplings of CFT primaries  \cite{Cardy,Kraus:2016nwo,Maloney,MMTwo}. 
 It is also not clear what should be the partition function of pure Einstein gravity, though there have been a number of attempts \cite{MW,Keller:2014xba,BCM}.
 
 Here we will consider a simpler problem.   We consider 2d CFT's with left and right central charges $(c_\ell,c_r)=(\D,\D)$ (for some positive
 integer $\D$) that also have left- and right-moving $\U(1)^\D$ current algebras.   It is expected that any such theory is in the family
 originally constructed by Narain \cite{Narain,NSW}, the parameter space being the locally symmetric space \be\label{wolllo}
 \M_\D=\SO(\D,\D;\Z)\backslash \SO(\D,\D;\R)/\SO(\D)\times \SO(\D).\ee
 (Here $\SO(\D,\D;\Z)$ must be understood as the automorphism group of an even integer unimodular lattice $\Lambda$ of signature $(\D,\D)$.)   
 As a CFT
 moduli space, $\M_\D$ carries a natural Zamolodchikov metric, which determines a natural measure.   This is actually the same
 metric and measure that $\M_\D$ gets because it is  locally homogeneous, that is, it is the quotient of the homogeneous space
 $ \SO(\D,\D;\R)/\SO(\D)\times \SO(\D)$ by the discrete group $\SO(\D,\D;\Z)$.   $\M_\D$ has finite measure for any $\D>1$, and when this is the case,
 it makes sense to average over $\M_\D$ in its natural measure.   This is what we will mean by ``averaging over Narain moduli space.''   
 %(The specific function whose average we want to take does not behave well enough at infinity for the average to make sense for $\D=2$; it is well-defined for any $\D>2$.)
 
 For a point $m\in\M_\D$, let $Z_\Sigma(m,\tau)$ be the partition function of the corresponding CFT on a Riemann surface $\Sigma$ with modular parameters $\tau$.
 The lattice sum that  controls the $m$-dependence of $Z_\Sigma(m,\tau)$ 
 is a nonholomorphic theta function that was originally introduced by C. L. Siegel and rediscovered by Narain; 
 we will call it the Siegel-Narain theta function, and denote it as\footnote{$\Theta(m,\tau)$
 is not holomorphic or anti-holomorphic 
 in $\tau$.   To emphasize this, one could denote it as $\Theta(m,\tau,\bar\tau)$.   To lighten notation, we will not do that.}  $\Theta(m,\tau)$.
 It turns out that the average of $\Theta(m,\tau)$  over $m\in \M_\D$  can be computed in a simple way,
 using what is known in number theory as the Siegel-Weil formula,  developed by Siegel, Maass, and Weil  \cite{Siegel,Maass,Weil,WeilTwo}.   
   The Siegel-Weil formula expresses the average over $m$ of $\Theta(m,\tau)$ in terms of a non-holomorphic
 Eisenstein series $E_{\D/2}(\tau)$ with modular weights $(\D/2,\D/2)$.\footnote{Non-holomorphic or real analytic Eisenstein series  may be less familiar than  holomorphic
 ones; however, they appear in the effective action of string theory \cite{GreenGutperle,GreenSethi}.  The constant term in the Siegel-Weil formula, which is known as the  Smith-Minkowski-Siegel mass formula, computes volumes of moduli spaces and has appeared studies of the moduli space of conformal field theories \cite{Ashok:2003gk, Moore:2015bba}.}

 If the ensemble of Narain theories
  is dual to a theory of gravity, that theory is not going to be a conventional one.   First of all, since the CFT's considered have $\U(1)^{2\D}$ current
 algebra and in particular $\U(1)^{2\D}$ global symmetry,
 the bulk theory will have $\U(1)^{2\D}$ gauge symmetry.   The perturbative anomalies of the boundary current algebra become Chern-Simons couplings in the bulk theory.
 Those anomalies are controlled by the even integer unimodular lattice $\Lambda_{IJ}$, $I,J=1,\cdots,2\D$ that is used in constructing the CFT.   Thus at a minimum 
 we expect the bulk theory to have gauge fields $A^I$, $I=1,\cdots, 2\D$ of the group $\U(1)^{2\D}$ with Chern-Simons couplings.\footnote{A variant that we consider in section \ref{comparison} is that the gauge group is really $\R^{2D}$.}  On a three-manifold $Y$, the 
 Chern-Simons action is
 \be\label{pilk} I_\CS=\sum_{I,J}\frac{\Lambda_{IJ}}{2\pi}\int_Y A^I\wedge \d A^J. \ee  
 The Narain CFT's require that $\Sigma$ should be oriented (because the target space $B$-field plays an important role).  To define the Chern-Simons action, $Y$
 should be oriented, in such a way that along the boundary its orientation induces the orientation of $\Sigma=\partial Y$.   
 
 The action (\ref{pilk}) is diffeomorphism-invariant without any need for a metric tensor.   
 One may think that  to get the dual theory we want, we must add a metric tensor and
 a gravitational action.  But there are reasons to believe that this is not the case.   In the boundary theory, the stress
 tensor can be expressed in terms of the currents via the Sugawara construction:  
 if $T$ is the holomorphic stress tensor, and $J^a$, $a=1,\cdots,\D$ are the holomorphic currents, the
 formula is $T(z)=\sum_a :J^aJ^a(z):$.    What is the bulk dual of the Sugawara formula?   It is plausible that the bulk dual of the fact that the stress tensor
 is a function of the currents rather than being ``new'' is that we should not introduce in bulk a metric tensor that is independent of the gauge fields.  
 Thus we might hope that the bulk dual of the average over Narain moduli space is simply the gauge theory with action $I_\CS$, or at least, something
 more like this than a theory with a dynamical metric tensor.   This has to be supplemented
 with a recipe for what $Y$ should be.   The Chern-Simons gauge theory {\it per se} does not suggest any specific rule to sum over $Y$'s with fixed conformal boundary
 $\Sigma$; in fact, the Chern-Simons path integral  on a three-manifold $Y$  makes sense for any particular $Y$, and the Chern-Simons gauge theory does not come with any rationale for summing over choices of $Y$.

 \begin{figure}
 \begin{center}
   \includegraphics[width=3in]{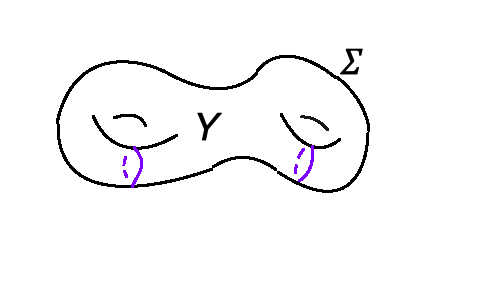}
 \end{center}
\caption{\small If a  closed Riemann surface $\Sigma$ is embedded in $\R^3$ in an arbitrary fashion, then its ``interior'' is, topologically,  a handlebody $Y$.   
Such an embedding of $\Sigma$ determines a distinguished sublattice $\Gamma_0$ of the first homology $\Gamma=H_1(\Sigma,\Z)$, spanned by one-cycles that
are contractible in $Y$. In the present example, $\Sigma$ has genus 2, and $\Gamma_0$ is spanned by the two one-cycles drawn.  \label{handlebody}}
\end{figure}
However, the Siegel-Weil formula suggests how to sum over $Y$ as long as $\Sigma$ is connected.   Topologically, the simplest class of three-manifolds
with boundary $\Sigma$ are ``handlebodies.'' An  orientable two-manifold $\Sigma$ can be embedded in $\R^3$ (in many topologically distinct ways).
Any such embedding divides the complement of $\Sigma$ in $\R^3$ into two components; the ``interior'' component is called a handlebody (fig. \ref{handlebody}).   
 With a plausible
recipe for how to compute $Z_Y(\tau)$ for $Y$ a handlebody, the Siegel-Weil formula gives
\be\label{woddo}\langle Z_\Sigma(m,\tau)\rangle =\sum_{Y\in\J} Z_Y(\tau),\ee
where $\langle ~~\rangle$ represents an average over $\M_\D$, and $\J$ is the set of all handlebodies with boundary $\Sigma$.

In pure Einstein gravity with negative cosmological constant, if $\Sigma$ has genus 1,
there is a semi-plausible justification for summing only over handlebodies \cite{MW}.  Any three-manifold that is a classical solution of Einstein gravity with negative
cosmological constant and that has  a
conformal boundary consisting of a single component of genus 1 is a handlebody.   Therefore, if the path integral of Euclidean quantum gravity should be constructed as an
expansion around critical points (or as an integral over Lefschetz thimbles associated to critical points), then only handlebodies can contribute if $\Sigma$ has genus 1.
(In a supersymmetric extension of three-dimensional gravity, there can be a more clear-cut justification to sum only over handlebodies in evaluating an appropriate
index  \cite{Farey}.) 

This justification to include only handlebodies is
 not entirely convincing for pure Einstein gravity, but in any case, we are here not considering pure Einstein gravity, but a more exotic theory that can
be approximated by the Chern-Simons theory (\ref{pilk}).   Moreover, even in pure  Einstein gravity, for $\Sigma$ of genus $g>1$, there are classical solutions
with conformal boundary $Y$ that are not handlebodies; see  \cite{Yin}.  

We can get, in a sense, a clearer picture of what the sum in eqn. (\ref{woddo}) would have to mean by considering the case that $\Sigma$ is not connected.
If $\Sigma$ is the disjoint union of components $\Sigma_\alpha$, $\alpha=1,\cdots, s$, with modular parameters $\tau_\alpha$, then for fixed $m\in\M_\D$, the partition function is a simple product: $Z_\Sigma(m,\tau)=\prod_\alpha
Z_{\Sigma_\alpha}(m,\tau_\alpha)$ (here we write $\tau$ for the whole collection of all the $\tau_\alpha$).    After averaging over $m$, this is of course no longer true:
\be\label{noddo} \langle Z_\Sigma(m,\tau)\rangle\not= \prod_\alpha \langle Z_{\Sigma_\alpha}(m,\tau_\alpha) \rangle. \ee

The Siegel-Weil formula gives an answer for $ \langle Z_\Sigma(m,\tau)\rangle$ also when $\Sigma$ is not connected, but to describe it, we have to first restate the
formula in the connected case.   Let $\Sigma$ be a connected Riemann surface of genus $g$.   The first homology group
$H_1(\Sigma,\Z)$ is a lattice  $\Gamma\cong \Z^{2g}$.   For $\gamma$, $\gamma'\in H_1(\Sigma,\Z)$, we denote their oriented intersection number as
$\la \gamma,\gamma'\ra$.   The pairing $\la~,~\ra$ is antisymmetric and nondegenerate.   A ``Lagrangian sublattice'' $\Gamma_0\subset H_1(\Sigma,\Z)$ is a primitive\footnote{\label{full} Here ``primitive''  
means that if $\Gamma_0$ contains a nonzero multiple of some $x\in H_1(\Sigma,\Z)$, then it actually contains $x$.  In other words, we are not allowed
to replace $\Gamma_0$ by a proper sublattice of the same rank.     
An equivalent definition is that a sublattice $\Gamma_0\subset H_1(\Sigma,\Z)$ is primitive if
the quotient $H_1(\Sigma,\Z)/\Gamma_0$ is torsion-free.  All sublattices considered in this paper are
assumed to be primitive; this condition is not always stated.}
sublattice of rank $g$ on which the intersection pairing vanishes; in other words, $\la\gamma,\gamma'\ra=0$ for $\gamma,\gamma'\subset \Gamma_0$.
For example, any set of $A$-cycles on $\Sigma$ determines a Lagrangian sublattice, namely the sublattice of $H_1(\Sigma,\Z)$ spanned by the homology classes of
those   $A$-cycles.   Alternatively, any oriented manifold $Y$ with boundary $\Sigma$ determines a Lagrangian sublattice, namely the sublattice  $\Gamma_0\subset H_1(\Sigma,\Z)$
that is spanned by all one-cycles $\gamma\subset \Sigma$ that are boundaries of two-manifolds in $Y$.   

If $\Sigma$ is connected, then every Lagrangian sublattice $\Gamma_0$  is associated in this way to a distinguished three-manifold $Y$, namely a handlebody.  
(To construct $Y$,
pick a set of $A$-cycles or disjoint simple closed curves in $\Sigma$ that provides a basis of $\Gamma_0$, and embed $\Sigma$ in $\R^3$ so that these $A$-cycles are contractible in the interior.)  
This gives a one-to-one correspondence between handlebodies and associated Lagrangian sublattices.   However, 
 associated to the same sublattice there are infinitely many other three-manifolds that are not handlebodies.  

For connected $\Sigma$, using the correspondence between handlebodies and Lagrangian sublattices of $H_1(\Sigma,\Z)$, 
we can reinterpret $\J$ as the set of  Lagrangian sublattices.    Once this is done, the Siegel-Weil formula says that
(with a plausible interpretation of $Z_Y(\tau)$) the formula (\ref{woddo}) holds for all $\Sigma$, connected or not.     The difference is that if $\Sigma$ is not
connected, there is no distinguished choice of a manifold $Y$ associated with a given Lagrangian sublattice.  So to state the formula in a way that is valid
whether $\Sigma$ is connected or not, we have to interpret
the sum on the right hand side of eqn. (\ref{woddo}) as a sum over Lagrangian sublattices.     We write it as such:
\be\label{nodo}\langle Z_\Sigma(m,\tau)\rangle =\sum_{\Gamma_0} Z_{\Gamma_0}(\tau),\ee
where the sum now runs over the set of Lagrangian sublattices and $Z_{\Gamma_0}(\tau)$ is the contribution to the partition function associated
to the Lagrangian sublattice $\Gamma_0$.

This tells us something about the exotic gravitational theory that is dual to an average over Narain moduli space -- if such a theory exists.   In this theory,
the objects that are analogous to ``manifolds with boundary $\Sigma$'' in ordinary geometry are classified by Lagrangian sublattices of $H_1(\Sigma,\Z)$.   
Apparently, the additional topological invariants that would be present in ordinary geometry are lacking in this more exotic theory.    

A conceivable alternative
interpretation might be that in the exotic theory, there are many ``manifolds'' associated to a given Lagrangian sublattice $\Gamma_0$, and the sum over all of these
is giving what we call $Z_{\Gamma_0}(\tau)$.  This seems less plausible, in part for the following reason.   
We will see that to reproduce the effect of averaging over $\M_\D$,
we have to use $Z_{\Gamma_0}(\tau)=c_{\Gamma_0}^\D$, where  $c_{\Gamma_0}$ depends on $\Gamma_0$ and $\tau$  but not on $\D$. 
If $Z_{\Gamma_0}(\tau)$ is built in a nontrivial way as a sum of contributions of different
``manifolds'' $\Gamma_{0,i}$,  then, since a nontrivial  formula $c_{\Gamma_0}^\D=\sum_i c_{\Gamma_{0,i}}^\D$ is not going to hold for all $\D$ no matter
what we assume for the  $c_{\Gamma_{0,i}}$,  individual contributions must have a more complicated dependence on $\D$, which then cancels out in the sum 
$\sum_i c_{\Gamma_{0,i}}^\D$.   It
seems more economical to assume that in the exotic theory under discussion, the analog of a ``manifold with boundary $\Sigma$'' 
is just classified by the choice of $\Gamma_0$.     This possibility does not sound completely far-fetched; it would be in keeping with the idea that among all the topological
invariants of classical geometry, only some that are particularly robust are well-defined in quantum gravity.

A close cousin of the $\U(1)^{2\D}$ current algebra considered in the present paper is a rational conformal field theory (RCFT), with an extended chiral
algebra that has only finitely many modules.  Attempts have been made to express RCFT partition functions in terms of sums over modular images of a  function which would be hypothetically the partition function of an exotic theory of gravity on a handlebody
 \cite{MGD,MoreEx}.  This program is successful  at $c=1/2$ at least in the sense that a suitable function exists, 
 except that one has to assume slightly puzzling equivalences between different handlebodies.   
 The program does not seem to work in the same way for theories  with  $1/2<c<1$.  Those theories  have a
chiral algebra that admits more than one modular-invariant partition function.   It may be that the bulk path integral in these cases represents a sort of average over
the possible boundary theories, given a knowledge of the chiral algebra.   
That is in the spirit of what we find here for a boundary theory with $\U(1)^\D$
left-moving and right-moving current algebras.   
The difference is that instead of finitely many possible boundary theories with a given chiral algebra,
we will have a continuous family of possible boundary theories, parametrized by $\M_\D$.

The organization of this article is as follows.
Section \ref{genusone} is devoted to averaging over Narain moduli space in the case of a surface $\Sigma$ of genus 1.  The Siegel-Weil formula that
carries out this averaging is explained in sections \ref{practice} and \ref{dtwo}.   In section \ref{gravint}, we attempt to interpret the output of the Siegel-Weil
formula in terms of a bulk dual theory.   In section \ref{adding}, we consider  what happens if we supply more information about the CFT by for example
specifying a particular extension of the boundary current algebra. This leads to a restricted averaging problem
that is governed by a more general version of the Siegel-Weil formula. 
In the most extreme case the average is a sum over a finite set of CFTs, rather than an integral over a moduli space; nevertheless, the result has a plausible gravity interpretation.

  In section \ref{highg}, we study the generalization to surfaces
of higher genus and to the case of a surface with multiple connected components.  The Siegel-Weil formula is applicable in any genus.
 As we have already explained, to interpret the result in terms of a bulk dual theory, we seemingly have to assume that the dual theory has a nonclassical
 notion of ``manifold with boundary
$\Sigma$.''  The Siegel-Weil formula also applies to the case of a surface $\Sigma$ with 
multiple connected components.   Connected correlators between different components would come, in ordinary gravity, from connected manifolds with
disconnected boundary, as in the case of JT gravity \cite{SSS}.   We study explicitly one example of a contribution to  a connected correlator,  which is related to the spectral  form factor $\langle Z_{T^2}(m,\tau') Z_{T^2}(m, \tau'') \rangle$.   The  contribution
that we analyze vanishes exponentially for large $\D$,
compared to the corresponding disconnected correlator.    This is as if adding a wormhole that connects two components
increases the classical action, which is probably the general state of affairs of a hypothetical theory of gravity related to the Siegel-Weil formula.  
The contribution to the spectral form factor we analyze is non-zero at large Lorentzian time, a feature which is indicative of theories with a discrete spectrum.
We discuss in section \ref{genuszero} the special case that $\Sigma$ is of genus 0 or has a component of genus 0.   
In averaging over Narain moduli space, a surface of genus 0 has no connected
correlator with anything else.   In the dual theory,  this might mean that there is no way to make a wormhole connecting a surface of genus zero
to another surface.   An alternative interpretation might be that in the dual theory, there is no notion of whether spacetime is connected.

Finally, in section \ref{gaugetheory}, we describe in more detail the attempt to interpret the dual of the average over Narain moduli space in terms of a gauge
theory with gauge group $\U(1)^{2\D}$ or possibly $\R^{2\D}$.     We find that when the boundary has a single connected component, bulk Chern-Simons theory on a handlebody, under some assumptions, exactly reproduces the corresponding term in the Siegel-Weil formula.  Indeed, there is a sense in which the bulk partition function is one-loop exact in a perturbative expansion in $1/\D$.\footnote{This is much simpler than the case of pure gravity, where the perturbative expansion is one-loop exact when the boundary is a torus \cite{MW} but is not expected to terminate when the boundary has higher genus \cite{Headrick:2015gba}.} However, we will discover that the situation is considerably more subtle when the boundary is disconnected.

 Appendix \ref{diffeq} contains further details about Narain moduli space, as well as a sketch of the 
 derivation of the Siegel-Weil formula at genus $g>1$ and $D>1$. 
 
 When this draft was largely complete, we learned of parallel work by Afkhami-Jeddi, Cohn, Hartman, and Tajdini \cite{ACHT}.

\section{The Siegel-Weil Formula In Genus One}\label{genusone}

\subsection{A Practice Case: $\D=1$}\label{practice}

We begin by discussing the Siegel-Weil formula in genus 1, where we can write somewhat more explicit formulas and the hypothetical gravitational dual
can be analyzed in a more direct way.   

In general, the Narain models are sigma-models with target a $\D$-torus $T^\D$, leading to a CFT with $(c_\ell,c_r)=(\D,\D)$.   We will begin with the case $\D=1$,
to illustrate the main idea.   We take the target space to be a circle of circumference $2\pi R$.
 The moduli space $\M_1$ in this example is parametrized by $R$,
which can be restricted to the range $1\leq R<\infty$ because of the $R\to 1/R$ duality symmetry.     Parametrizing the circle by an angle-valued field $X$,
the action (with conventions as in \cite{Polchinski}, eqn. (2.1.1)) is
\be\label{bonoc} I=\frac{R^2}{4\pi \alpha'} \int\d^2\sigma ~\partial_\alpha X\partial^\alpha X. \ee
The marginal operator associated with a deformation $R\to R+\delta R$ is $\frac{R\delta R}{2\pi\alpha'} \partial_\alpha X\partial^\alpha X$, and its two-point function
is proportional to $(\delta R)^2/R^2$.   This gives the Zamolodchikov metric of $\M_1$
\be\label{jugu} \d s^2=4\frac{\d R^2}{R^2},\ee
which is also the natural metric on $\M_1$ as a locally homogeneous space.  (The factor of 4 is chosen for later convenience and to agree with the more
general formula \ref{zamet} of Appendix \ref{diffeq}.)

The partition function of the model on a genus 1 surface $\Sigma$ with modular parameter $\tau=\tau_1+\i \tau_2$ is (see \cite{Polchinski}, eqn. (8.2.9))
\be\label{zacko}Z_\Sigma(R,\tau)= \frac{\Theta(R,\tau)}{|\eta(\tau)|^2},\ee
where $\eta(\tau)$ is the Dedekind eta function 
\be\label{nacko}\eta(\tau)=q^{1/24}\prod_{n=1}^\infty(1-q^n),~~~~~q=\exp(2\pi \i\tau),  \ee
and $\Theta(R,\tau)$ is the $\D=1$ case of the Siegel-Narain theta function,
\be\label{sn} \Theta(R,\tau)=\sum_{n,w\in\Z}Q(n,w;R,\tau),\ee
with
\be\label{thetaf} Q(n,w;R,\tau)=\exp\left(-\pi\tau_2\left(\frac{\alpha' n^2}{R^2}+\frac{w^2R^2}{\alpha'}\right)+2\pi\i
\tau_1 nw\right). \ee
The integers $n$ and $w$ are the momentum and winding quantum numbers of a string. 

A small calculation gives   
\begin{align} \tau_2\frac{\partial}{\partial\tau_2} Q&=-\pi\tau_2\left(\frac{\alpha' n^2}{R^2}+\frac{w^2R^2}{\alpha'}\right) Q\cr
 \tau_2^2\frac{\partial^2}{\partial\tau_2^2}Q&=\pi^2\tau_2^2\left( \frac{\alpha' n^2}{R^2}+\frac{w^2R^2}{\alpha'}\right)^2 Q  \cr
 \tau_2^2\frac{\partial^2}{\partial\tau_1^2} Q & = -4\pi^2\tau_2^2n^2w^2 \cr
 \left(R\frac{\partial}{\partial R}\right)^2 Q&  = \left(4\pi^2\tau_2^2\left( \frac{\alpha' n^2}{R^2}-\frac{w^2R^2}{\alpha'}\right)^2-4\pi\tau_2\left(\frac{\alpha' n^2}{R^2}+\frac{w^2R^2}{\alpha'}\right) \right) Q\,.  \end{align}
Therefore
\be\label{fireq} \left(\tau_2^2\left(\frac{\partial^2}{\partial\tau_1^2}+\frac{\partial^2}{\partial \tau_2^2}\right) +\tau_2\frac{\partial}{\partial \tau_2} -\frac{1}{4}\left(R
\frac{\partial}{\partial R}\right)^2\right) Q = 0, \ee
and it follows immediately that $\Theta$ obeys the same equation:
\be\label{nireq} \left(\tau_2^2\left(\frac{\partial^2}{\partial\tau_1^2}+\frac{\partial^2}{\partial \tau_2^2}\right) +\tau_2\frac{\partial}{\partial \tau_2} -\frac{1}{4}\left(R
\frac{\partial}{\partial R}\right)^2\right) \Theta(R,\tau) = 0\,. \ee

The measure on $\M_1$ can be deduced from the metric (\ref{jugu}) and is
\be\label{worfo}\mu(R)=\frac{\d R}{2R}. \ee
The volume of $\M_1$ is infinite, so averaging over $\M_1$ does not make sense, as remarked in the introduction.
Let us ignore this for a moment and explain the strategy in the derivation of the Siegel-Weil formula.    We would define
a function $F_1(\tau)$ by integrating $\Theta(R,\tau)$ over $\M_1$:
\be\label{yuf} F_1(\tau)=\int_1^\infty\frac{\d R}{2R}  \Theta(R,\tau). \ee
In reality, this integral does not converge, since $\Theta\sim R$ for $R\to\infty$.    Then using the differential equation for $\Theta$, we 
find that
\be\label{uf} \left(\tau_2^2\left(\frac{\partial^2}{\partial\tau_1^2}+\frac{\partial^2}{\partial \tau_2^2}\right) +\tau_2\frac{\partial}{\partial \tau_2}\right)F_1(\tau)
= \frac{1}{8}\int_1^\infty \frac{\d R}{R} \left(R
\frac{\partial}{\partial R}\right)^2 \Theta(R,\tau)=\frac{1}{8}\int_1^\infty \d R \frac{\partial}{\partial R} \left(R \frac{\partial \Theta(R,\tau)}{\partial R}\right). \ee
Next one tries to integrate by parts to prove the vanishing of the right hand side of this equation.  There
is no surface term at $R=1$ because the $R\to 1/R$ symmetry ensures that $\partial_R\Theta|_{R=1}=0$.  If there were also no surface term
at $R=\infty$, we would deduce a differential equation for $F_1(\tau)$: \be\label{nuf} 
 \left(\tau_2^2\left(\frac{\partial^2}{\partial\tau_1^2}+\frac{\partial^2}{\partial \tau_2^2}\right) +\tau_2\frac{\partial}{\partial \tau_2}\right)F_1(\tau)=0.\ee
The derivation of the Siegel-Weil formula would be completed by using this differential equation together with a knowledge of the behavior for $\tau_2\to\infty$ 
to determine $F_1(\tau)$. 

The only problem with this derivation is that the  behavior for $R\to\infty$  does not allow either the definition of $F_1(\tau)$ in eqn. (\ref{yuf}) or the integration
by parts that would show the vanishing of  eqn. (\ref{uf}).    Hence we will move on to the case of larger $\D$, where such a derivation does work.

Before doing so, we restate the differential equation in a convenient form.  
The Laplacian of $\M_1$  in the metric  (\ref{jugu}) is
\be\label{wuggu}\Delta_{\M_1}=-\frac{1}{4}\left( R\frac{\d}{\d R}\right)^2. \ee
The natural metric of the upper half plane $\H$ is 
\be\label{ugu} \d s^2=\frac{\d\tau_1^2+\d\tau_2^2}{\tau_2^2}. \ee
Acting on a scalar function, the Laplacian of the upper half plane is
\be\label{nugu} \Delta_\H=-\tau_2^2\left(\frac{\partial^2}{\partial\tau_1^2}+\frac{\partial^2}{\partial\tau_2^2}\right). \ee
Therefore the differential equation for the Siegel-Narain theta function for $\D=1$ can be written
\be\label{plugu} \left(\Delta_\H-\tau_2\frac{\partial}{\partial \tau_2}-   \Delta_{\M_1}\right)\Theta(R,\tau)=0. \ee 

\subsection{The Siegel-Weil Formula for Higher $\D$}\label{dtwo}

For general $\D$, we consider a sigma-model with target $T^\D$ and general (constant) metric $G$ and two-form field $B$.    $G$ and $B$ together are
the moduli that parametrize $\M_\D$; we schematically denote these moduli as $m$.    For $\D>1$, $\M_\D$ has finite volume\footnote{ 
For example, for $\D=2$, using the relationships $\SO(2,2,\R)\cong (\SL(2,\R)\times \SL(2,\R))/\Z_2$ and $\SO(2,2,\Z)\cong (\SL(2,\Z)\times \SL(2,\Z))/\Z_2$, one can show that $\M_2$ is the product of two copies of $\SL(2,\Z)\backslash\H$, with 
$\H = \SL(2,\R)/U(1)$ being the upper half plane.  $\SL(2,\Z)\backslash \H$ is isomorphic to the moduli space of Riemann surfaces of genus 1 and
has finite volume.  
Using the metric \ref{zamet}, one can show that as  $\D$ increases,  the volume integral converges more rapidly.   The same
is true for the integral in (\ref{defav}). To verify the last statement, one has to take into account the behavior of the function $\Theta$ near infinity in the space of target
space metrics.}  and we normalize its measure $\mu(m)$  so that the volume is 1:
\be\label{wytto} \int_{\M_\D}\d\mu(m)=1. \ee

The partition function of the sigma-model in genus 1 is
\be\label{wackor}Z_\Sigma(m,\tau)= \frac{\Theta(m,\tau)}{|\eta(\tau)|^{2\D}},\ee
where $\Theta(m,\tau)$ is the Siegel-Narain theta function that comes from a sum over momenta and windings.  See eqn.  \ref{kolo} for the
explicit formula.   

$\Theta(m,\tau)$ obeys a differential equation that generalizes eqn. (\ref{plugu}) for $d=1$:
\be\label{zugu}  \left(\Delta_\H-\D\tau_2\frac{\partial}{\partial \tau_2}- \Delta_{\M_1}\right)\Theta(R,\tau)=0. \ee 
A derivation of this equation similar to the one in section \ref{practice} is sketched in Appendix \ref{diffeq}. 

The next step is to average $\Theta(m,\tau)$ over the Narain moduli space $\M_\D$, defining
\be\label{defav}F_\D(\tau)=\int_{\M_\D}  \d\mu(m)\,\Theta(m,\tau). \ee
Actually, this integral converges only for $\D>2$, so in continuing, we make that restriction.   ($\M_2$ has finite volume, but $\Theta(m,\tau)$ grows
at infinity 
in such a way that the integral is divergent for $\D=2$.  To be precise, $\Theta(m,\tau)$ grows in the limit that the target space volume is large and also in the
``large complex structure'' limit.)    
Following the same steps as in section \ref{practice}, we deduce from the last two
formulas a differential equation
for $F_\D(\tau)$:
\be\label{drofo} \left(\Delta_\H-\D\tau_2\frac{\partial}{\partial\tau_2}\right) F_\D(\tau)= 0 . \ee 

In addition to satisfying this differential equation, $F_\D(\tau)$ transforms under modular transformations with
weights\footnote{To say that a function $f$ has modular weights $(u,v)$  with $u-v\in \Z$ means that $f((a\tau+b)/(c\tau+d)) =(c\tau+d)^u(c\bar\tau+d)^v f(\tau)$ for
$\begin{pmatrix} a& b \cr c & d\end{pmatrix}\in \SL(2,\Z)$.     For
example, $\tau_2=\mathrm{Im}\,\tau$ has modular weights $(-1,-1)$. The function $|\eta(\tau)|^2$ has modular weights $(1/2,1/2)$, so $F_\D(\tau)$ must have
 modular weights $(\D/2,\D/2)$ to ensure modular invariance of the partition function.  The case $u-v\in\frac{1}{2}+\Z$ is more complicated and will appear later.}       
 $(\D/2,\D/2)$, since $F_\D(\tau)/|\eta(\tau)|^{2\D}$ is modular-invariant.  In addition, 
\be\label{rofo}\lim_{\tau_2\to\infty} F_\D(\tau)=1, \ee
since  $\lim_{\tau_2\to\infty}\Theta(m,\tau)=1$. 

It is convenient to define
$ \W_\D(\tau)=\tau_2^{\D/2} F_\D(\tau)$.   This function is modular-invariant, since multiplying by $\tau_2^{\D/2}$
cancels the modular weights of $F_\D(\tau)$.   Clearly $ \W_\D(\tau)\sim\tau_2^{\D/2}$ for $\tau_2\to\infty$.
Finally the differential equation for $F_\D(\tau)$ becomes
\be\label{difg}\left(\Delta_\H +s(s-1)\right)  \W_\D(\tau)=0, ~~~~ s=\D/2. \ee
Thus $ \W_\D(\tau)$ is an eigenfunction of $\Delta_\H$ with the eigenvalue $-s(s-1)$, which is negative for $\D>2$.   

For $\D> 2$, the  differential equation (\ref{difg}) has no nonzero solution that grows at infinity more slowly than $\tau_2^{\D/2}$.   This fact was important
in \cite{GreenGutperle,GreenSethi}.   Indeed, any solution of the differential equation that grows more slowly than $\tau_2^{\D/2}$ is bounded by a constant
times $\tau_2^{1-\D/2}$, and therefore (for $\D>2$) is a square-integrable eigenfunction of the Laplacian with  the negative eigenvalue $-s(s-1)$.   But the Laplacian on
any manifold, acting on square-integrable wavefunctions,
is strictly non-negative.

A function that satisfies all of the necessary conditions is the non-holomorphic (real analytic)  Eisenstein series\footnote{{It is important to note that this real analytic Eisenstein series is different from the holomorphic Eisenstein series (which transforms with modular weight $(n,0)$) that commonly appears in the theory of modular forms.}}
\be\label{nifg} E_s(\tau) =\sum_{(c,d)=1}  \frac{\tau_2^s}{|c\tau+d|^{2s} }.\ee
The sum is over pairs of  relatively prime integers $c,d$, up to sign (that is, we do not distinguish $(-c,-d)$ from $(c,d)$).   
Alternatively, the sum is over all modular images of the function $\tau_2^s$, since
a general element $\begin{pmatrix}a&b\cr c&d\end{pmatrix}\in\SL(2,\Z)$ maps $\tau_2^s$ to $\tau_2^s/|c\tau+d|^{2s}$.  
So the sum in equation (\ref{nifg}) can be alternatively written as 
\be\label{eismod}
 E_s(\tau) =\sum_{\gamma \in  P \backslash SL(2,\Z)}  \Im\, \left(\gamma \tau\right)
 \ee
where $ P  = \left\{ \left({1~n\atop 0~1}\right)\right\}$ is the subgroup of $SL(2,\Z)$, isomorphic to $\Z$, that leaves ${\rm Im}\,\tau$ invariant.
It is straightforward to check that the coprime integers $(c,d)$ uniquely label elements of the coset $ P \backslash SL(2,\Z)$.
The sum in eqn. (\ref{nifg}) converges
for $\mathrm{Re}\,s>1$, which in our application means $\D>2$.    

 Since $E_s(\tau)$ is a sum over
all of the modular images of $\tau_2^s$, it is modular-invariant.    The function $\tau_2^s$ is easily seen to be an eigenfunction of $\Delta_\H$ with
eigenvalue $-s(s-1)$; the same therefore is true of its modular images, and of $E_s(\tau)$.   Finally, it is immediate that $E_s(\tau)\sim \tau_2^s$ for $\tau_2\to\infty$.
Thus, $E_{D/2}(\tau)$ satisfies all of the desired properties of $ \W_\D(\tau)$.    These functions must be equal, since  their difference $ \W_\D(\tau)-
E_{\D/2}(\tau)$ grows at infinity more slowly than $\tau_2^{\D/2}$ and hence must vanish, as discussed earlier.  

Finally, we get an explicit formula for the average of the  genus 1 partition function over the Narain moduli space $\M_\D$:
\be\label{hubb}\langle Z_\Sigma(m,\tau)\rangle =\frac{E_{\D/2}(\tau)}{\tau_2^D|\eta(\tau)|^{2\D}}.\ee
The numerator and denominator are both modular-invariant.   

\subsection{Gravitational Interpretation Of The Formula}\label{gravint}

Our next task is to provide a possible interpretation of this formula in terms of an exotic bulk theory that is dual to an average over Narain moduli space.

As discussed in the introduction, the starting point is to assume that the bulk partition function, for the case that the conformal boundary is a surface $\Sigma$
of genus 1, should be expressed as a sum over handlebodies.    Let us decompose the genus 1 surface $\Sigma$ as $S^1\times S^1$, where the first
factor parametrizes ``space,'' and the second factor parametrizes ``Euclidean time.''    One particular handlebody $Y$ with boundary $\Sigma$ is obtained by
filling in the first factor by a two-dimensional disc $ \dD_2$.   Thus $Y\cong  \dD_2\times S^1$.    
This handlebody can be obtained by Wick rotating Lorentzian AdS$_3$ to Euclidean time via $t\to i t_E$, where $t$ is the usual global time coordinate, and then periodically identifying $t_E$. 
This handlebody is usually referred to as thermal AdS, since it is the one used to study thermal physics in an AdS background.

Any other handlebody with boundary $\Sigma$ is obtained
from $Y$ %this one 
by a modular transformation of the boundary.  In other words, to construct a more general handlebody one takes some other decomposition
of $\Sigma$ as $S^1\times S^1$, and fills in the first factor by a disc.    These other handlebodies are thus labelled by elements of the modular group $SL(2,\Z)$.
In fact, because the element $T^n=\left({1~n\atop 0~1}\right)\in SL(2,\Z)$ does not generate a new handlebody, each handlebody is uniquely labelled by an element of the coset $ P \backslash SL(2,\Z)$, where $ P $ is the subgroup of triangular matrices generated
by $T$.  
One simple example is the handlebody obtained from thermal AdS by an $S$ transform -- this handlebody is obtained by filling in the ``Euclidean time" circle, rather than the spatial circle.  This handlebody is the Euclidean continuation of the BTZ black hole in AdS.

In \cite{MW}, a Hamiltonian approach was used to evaluate the path integral of Einstein gravity on $ \dD\times S^1$.   In this approach, the key step is to determine
the spectrum of physical states that arise in quantization on the spatial manifold $ \dD$.  The partition function on $ \dD\times S^1$ is then evaluated as a trace in
that Hilbert space.    In Einstein gravity, there are no bulk excitations; the only physical
states in quantization on $ \dD$ are the ``boundary gravitons,'' first described by Brown and Henneaux \cite{BH}.   The proposal in \cite{MW} was that the path
integral on $ \dD\times S^1$ simply equals the partition function of the Brown-Henneaux modes.  
In other words, $\dD\times S^1$ is thermal AdS, and in three spacetime dimensions, the only excitations in thermal AdS
are the Brown-Henneaux modes.
The resulting formula for the gravitational path integral on $\dD\times S^1$
 was later confirmed by a direct 1-loop computation\footnote{There
can be no higher order corrections, since the energy and momentum of the boundary gravitons are uniquely determined by conformal invariance along the
boundary.} 
in Einstein gravity \cite{GMY}.

In the present context, instead of the boundary gravitons, we should discuss the boundary modes of the current algebra.   In other words, instead of
Einstein gravity, we are here considering a theory that is supposed to be approximated, in some sense, by the $\U(1)^{2\D}$ Chern-Simons theory 
 \be\label{pillk} I_\CS=\sum_{I,J}\frac{\Lambda_{IJ}}{2\pi}\int_Y A^I\wedge \d A^J, \ee
 where $\Lambda  $ is an even integral unimodular form of signature $(\D,\D)$. 
Instead of boundary gravitons, the Chern-Simon theory, if treated as in \cite{Witten}, has
 chiral and anti-chiral boundary current algebras   which are 
abelian (since in this case we are studying an abelian gauge theory)
and are each of rank $\D$ (because of the signature of the quadratic form).   As explained in the introduction, this relation of the bulk Chern-Simons theory
to the boundary current algebra was the rationale for introducing the Chern-Simons theory.    

The partition function of the boundary current algebras is the same as the partition function of $\D$ left- and right-moving chiral bosons, with zero-modes omitted.
 It is simply 
\be\label{ZCS} Z^{D\times S^1}_{CS} = \frac{1}{|\eta(\tau)|^{2\D}} = |q|^{-\D/12} \prod_{n=1}^\infty \frac{1}{|1-q^n|^{2\D}}. \ee   
This is  can be interpreted as the thermal partition function of a gas of $\D$ ``boundary photons," in the same way that the gravity partition function was the thermal partition function of a gas of boundary gravitons.  It is the vacuum character of $\D$ copies of the $U(1)\times U(1)$ current algebra.  This expression can be verified in a direct bulk computation in $U(1)^{2\D}$ Chern-Simons theory, just as in the gravity case. This will be discussed in section \ref{gaugetheory}. One important feature to note is that we have not included any separate factors of $|q|$ in equation (\ref{ZCS}), aside from the factors of $q^{1/24}$ which are contained in the definition of $\eta(\tau)$.  In a normal theory of gravity such a factor would come from the classical Einstein action of the saddle point.  In the present case we have not included a separate Einstein-Hilbert term in the action since, as explained in the introduction, the boundary stress tensor is itself an element of the $U(1)^{2\D}$ current algebra.  The factor of $|q|^{-\D/12}$ in equation (\ref{ZCS}) comes entirely from the bulk Chern-Simons computation, as we will discover in section \ref{gaugetheory}, and can be regarded as a one-loop contribution of bulk Chern-Simons theory to the effective cosmological constant.
We note also that equation (\ref{ZCS}) is one-loop exact because, as in the gravity case, the form of the answer is entirely fixed by the structure of the $U(1)^{2\D}$ current algebra.

Equation (\ref{ZCS}) is the result of the bulk path integral for one particular handlebody with boundary $\Sigma$.
To get the full partition function we need to sum over all handlebodies.  That is, we must compute
\be
Z^{bulk} = \sum_{\gamma \in  P  \backslash SL(2,\Z)} \frac{1}{|\eta(\gamma \tau)|^{2\D}}\,.
\ee  
This is a much more straightforward problem than the superficially similar problem that was treated in \cite{MW}.   We simply write
\be\label{judg}\frac{1}{|\eta(\tau)|^{2\D}}=\frac{1}{\tau_2^{\D/2}|\eta(\tau)|^{2\D}}\cdot \tau_2^{\D/2}. \ee
The function $\tau_2^\D|\eta(\tau)|^{2\D}$ is modular invariant.  So summing over modular images does nothing to this function.    Thus we just need the sum over modular
images of the function $\tau_2^{\D/2}$.     But this sum was already done in equation (\ref{eismod}); it equals the real-analytic Eisenstein series $E_{\D/2}(\tau)$.   

So given our assumptions, the result that comes from summing over handlebodies is
\be\label{wudg} Z^{bulk} = \frac{E_{\D/2}(\tau)}{\tau_2^{\D/2}|\eta(\tau)|^{2\D}}, \ee
which as we learned in section \ref{dtwo} is equal to the average of the partition function of the boundary CFT over the moduli space $\M_\D$.  

We consider the generalization of this derivation to higher genus in section \ref{highg}.

\begin{comment}
In \cite{GMY}, a detailed calculation was carried out in Einstein gravity to confirm the claim that the gravitational path integral on a genus one handlebody
reproduces the partition function of the Brown-Henneaux modes.    Using the same formulas, one can
 evaluate a gauge theory partition function on a genus one handlebody and  confirm, with some assumptions,
  our claim that it equals the partition function of the current algebra modes.
See section \ref{gaugetheory}.
\end{comment}

\subsection{Adding More Information About the CFT}\label{adding}

We have so far considered the Siegel-Weil formula only for the case that the lattice $\Lambda$ is even and unimodular, as well as integral.   In number theory,
this restriction would be considered slightly artificial; there is a Siegel-Weil formula for an arbitrary integer lattice.    Here we will sketch how this generalization can
arise in a variant of the problem that we have considered so far.   We will not be as detailed as we were in the case of an even unimodular lattice.   In particular,
we will not try to provide proofs of the more general version of the Siegel-Weil formula that we will invoke, though it appears that the approach in
Appendix \ref{diffeq}  can potentially be generalized.

Up to this point, we have considered a boundary CFT about which we know nothing except that it has central charges $(\D,\D)$ and left- and right-moving
$\U(1)^\D$ current algebras.   We did not assume any knowledge about the dimensions of primary fields of this theory.   Instead we averaged over all possibilities,
getting an answer with a plausible gravitational interpretation. 

We could instead input some knowledge about the spectrum of primary fields and average only over the remaining possibilities.    It turns out that this
leads to more general versions of the Siegel-Weil formula.  

As a special case, let us suppose that we know that the CFT has a primary field for the current algebra of dimension $(1,0)$.   The condition under which this
occurs is as follows.    The vector space $V=\Lambda\otimes_\Z \R$ has a  metric  of signature $(\D,\D)$ that comes from the intersection form on the lattice $\Lambda$
and  does not depend on the CFT moduli.
Once one specifies those moduli, the metric $G$ of the torus determines the dynamics of the CFT fluctuations and  one gets
a decomposition $V=V_+\oplus V_-$, where $V_+$ and $V_-$ are subspaces on which the intersection form is positive or negative definite and (with a suitable
orientation convention) are respectively the spaces of left- and right-moving modes of the CFT.   Generically, neither $V_+$ nor $V_-$
contain any points of the lattice $\Lambda$; both $V_+\cap\Lambda$ and $V_-\cap\Lambda$ are generically empty.
The condition for the CFT to have a current algebra primary of dimension $(1,0)$ is that there should be a point $x\in V_+\cap \Lambda$ of length squared
2 (it does not matter if this length is computed using the indefinite signature metric of $\Lambda$ or the positive-definite metric of $V_+$; these coincide
for vectors in $V_+$).    The existence of this $(1,0)$ primary, along with a second one that is associated to the vector $-x$, which also lies in $V_+\cap\Lambda$,
extends a $\U(1)$ subalgebra of the CFT current algebra to $\SU(2)$ at level 1, which we denote as $\SU(2)_1$.   
(If we want to specify the level of the $\U(1)^{2\D}$ current algebra, we could call it $\U(1)^{2\D}_\Lambda$ and then the extended current algebra would be
$\SU(2)_1\times \U(1)^{2\D-1}_{\Upsilon_\perp}$, where $\Upsilon_\perp$ is introduced momentarily.   We will not use this notation because it is not
clear that subtleties concerning the level of the abelian current algebra are meaningful in the present context.  See section \ref{comparison}.)

Specifying the existence of such an $x$ reduces the CFT moduli space from $\M_\D$ to a subspace $\M_{\D,x}$.   Roughly speaking,
$\M_{\D,x}=\SO(\D-1,\D;\Z)\backslash SO(\D-1,\D;\R)/\SO(\D-1)\times \SO(\D)$.    However, one has to clarify what is meant by $\SO(\D-1,\D;\Z)$.
Let $\Upsilon_0$ be the rank 1 sublattice of $\Lambda$ that is generated by $x$, and let $\Upsilon_\perp$ be its orthocomplement.   
The group $\SO(\D-1,\D;\Z)$ that appears in the definition of $\M_{\D,x}$ is the automorphism group of $\Upsilon_\perp$.   

$\Upsilon_0$ is even but
not unimodular; its quadratic form is the $1\times 1$ matrix 2, so its discriminant is the determinant of that matrix, or 2.   Likewise $\Upsilon_\perp$
has discriminant 2.   In particular $\Lambda$ is not the tensor product $\Upsilon_0\times \Upsilon_\perp$; $\Lambda$ has discriminant 1, and
$\Upsilon_0\times \Upsilon_\perp$ has discriminant $2\times 2=4$.   The relation between them is
\be\label{reln} \Lambda=\Upsilon_0\otimes \Upsilon_\perp \oplus \Upsilon', \ee
where $\Upsilon'$ is a coset of $\Upsilon_0\otimes\Upsilon_\perp$.   More specifically, if  $y\in \Lambda$ is any vector whose inner product with $x$ is an odd integer,
then $\Upsilon'$ consists of vectors of the form $y+z$, $z\in \Upsilon_0\otimes \Upsilon_\perp$.
The Siegel-Narain theta function has a corresponding decomposition
\be\label{beln}\Theta(m,\tau)=\Theta_{\Upsilon_0\otimes \Upsilon_\perp}+\Theta_{\Upsilon'}, \ee
where $\Theta_{\Upsilon_0\otimes \Upsilon_\perp}+\Theta_{\Upsilon'}$ are computed, respectively, by sums over lattice points in $\Upsilon_0\otimes \Upsilon_\perp$
and in $\Upsilon'$. 

If we consider left-moving modes to be holomorphic and right-moving ones antiholomorphic, then the momentum-winding sum of $\Lambda_0$ is
a holomorphic theta function
\be\label{worp}\theta(\tau)=\sum_{n\in\Z} q^{n^2},~~~ q=\exp(2\pi\i\tau). \ee
Because $\Lambda_0$ is not unimodular, this function is not mapped to itself by modular transformations.   Rather, 
\be\label{norp}\theta(-1/\tau)=\sqrt{\frac{\tau}{2\i}}(\theta(\tau)+\t\theta(\tau)),~~~\t\theta(\tau)=\sum_{r\in \Z+1/2} q^{r^2}. \ee
The functions $\theta$ and $\t\theta$ are associated to the two characters of the chiral algebra $\SU(2)_1$.

Let $m'$ be the CFT moduli that remain after we insist on the existence of the vector $x\in \Lambda\cap V_+$.   In other words,
$m'$ are the Narain moduli of $\Upsilon_\perp$.
The expansion of $\Theta(m',\tau)$ in terms of theta functions of $\Lambda_0$ and $\Lambda_\perp$ is
\be\label{wacko}\Theta(m',\tau)=\theta(\tau)\Theta_{\Upsilon_\perp}(m',\tau)+ \t\theta(\tau) \t\Theta_{\Upsilon_\perp}(m',\tau), \ee
where $\Theta_{\Upsilon_{\perp}}(m',\tau)$ is the momentum-winding sum of the lattice $\Upsilon_\perp$, and $\t\Theta_{\Upsilon_\perp}(m',\tau)$
is a second function into which this transforms under modular transformations.   The two terms on the right hand side of (\ref{wacko}) are
associated to the two summands in eqn. (\ref{reln}).   

Thus, in order to compute an average CFT partition function, we need to average $\Theta_{\Upsilon_\perp}(m',\tau)$ and $\t\Theta_{\Upsilon_\perp}(m',\tau)$ over
$\M_{\D,x}$.   We will write $\langle Z_\Sigma(m',\tau)\rangle_x$ for the average of $Z_\Sigma(m',\tau)$ over $\M_{\D,x}$.
 The Siegel-Narain formula for the lattice $\Upsilon_\perp$ expresses this average in terms of real-analytic Eisenstein series $E(\tau)$
and $\t E(\tau)$ of weights $(-1/2,0)$:
\be\label{acko}\langle Z_\Sigma(m',\tau)\rangle_x =\frac{ \theta(\tau)E(\tau)+\t\theta(\tau)\t E(\tau)}{\tau_2^{\D/2} |\eta(\tau)|^{2\D}}.\ee
The definition of a function of modular weight $(-1/2,0)$ (or more generally of modular weight $(u,v)$ with $u-v\in\frac{1}{2}+\Z$) is rather subtle.
The simplest definition is simply to say that $E(\tau)$ and $\t E(\tau)$ transform in such a way that the expression on the right hand side of eqn. (\ref{acko})
is modular invariant.   A definition rather along these lines (for holomorphic forms)
is given in Chapter IV of \cite{Koblitz}. $E(\tau)$ and $\t E(\tau)$ are given by formulas
similar to eqn. (\ref{nifg}), with an extra factor $1/(c\tau+d)^{1/2}$ in the denominator on the right hand side; there are also some congruence conditions on $c$ and $d$,
and one has to include some roots of unity in the sum to compensate for such factors in the modular transformations of the theta functions.   
Details are described in \cite{Koblitz} for the case of holomorphic modular forms of half-integral weight.  

Qualitatively, eqn. (\ref{acko}) is in agreement with what we might expect in a bulk analysis along the lines of section \ref{gravint}.   
As already noted, the  existence of a vector $x\in \Lambda\cap V_+$
extends the left-moving current algebra from $\U(1)^\D$ to $\SU(2)_1\times \U(1)^{\D-1}$.   The right-moving current algebra is still $\U(1)^\D$.
So the natural bulk Chern-Simons theory is $\SU(2)_1\times \U(1)^{2\D-1}$.    The starting point in trying to compute a bulk partition function
is to determine the partition function of a handlebody $D\times S^1$ by taking a trace in the Hilbert space associated with quantization on $D$.  In a hypothetical
bulk theory that can be approximated in some sense by Chern-Simons theory of $\SU(2)_1\times \U(1)^{2\D-1}$, the natural physical states in quantization
on $D$  are the current algebra modes on the boundary, and the corresponding partition function is $\theta(\tau)/|\eta(\tau)|^{2\D}$.   To derive this formula, one just needs to know that the partition function of a holomorphic or antiholomorphic
$\U(1)$ current algebra is $1/\eta(\tau)$ or $1/\bar{\eta(\tau)}$, while the partition function of the vacuum module of holomorphic $\SU(2)_1$ current algebra is
$\theta(\tau)/\eta(\tau)$.

To get an ansatz for the bulk partition function in this situation, we sum over modular images of  $\theta(\tau)/|\eta(\tau)|^{2\D}$.   Writing
\be\label{wondo}\frac{\theta(\tau)}{|\eta(\tau)|^{2\D}}= \frac{\theta(\tau)\tau_2^{\D/2}}{\tau_2^{\D/2}|\eta(\tau)|^{2\D}},\ee
where the denominator is modular-invariant, we see that have to sum over the modular images of $\theta(\tau)\tau_2^{\D/2}$.   This will generate
the numerator on the right hand side of eqn. (\ref{acko}).   The details are somewhat complicated because  the modular
transformation of $\theta(\tau)$ is somewhat complicated, so we will not attempt more detail.

We have considered the special case that a $\U(1)$ subgroup of the current algebra is extended to $\SU(2)$, but one can analyze in a similar way
any assumed extension of the $\U(1)^\D\times \U(1)^\D$ current algebra.\footnote{The extension does not necessarily involve an enhanced symmetry group.
For example, if we had assumed a vector $x\in \Lambda\cap V_+$ with $x^2=2r$, $r>1$, we would get $\U(1)$ current algebra at level $r$.}   In general, averaging over the remaining moduli via the Siegel-Weil formula
always gives a result that has a more or less plausible interpretation in terms of an exotic bulk theory of gravity.   We will just describe the construction that leads to the
holomorphic case of the Siegel-Weil formula.    Suppose that $\D$ is a multiple of 8, so that positive-definite even integer unimodular lattices  of rank $\D$
exist.   Let $\Lambda_-$ be such a lattice, and let us stipulate that the CFT moduli are such that
$\Lambda\cap V_-\cong \Lambda_-$.    This corresponds to a particular extension of
the right-moving $\U(1)^\D$ current algebra.   For example, if $\D=8$, there is only one choice for $\Lambda_-$, namely the ${\mathrm E}_8$ lattice, and $\U(1)^8$
is extended to ${\mathrm E}_8$ current algebra at level 1.   With our stipulation that $\Lambda\cap V_-\cong\Lambda_-$, it follows
that $\Lambda\cap V_+$ is equal to a possibly inequivalent even integer unimodular lattice $\Lambda_+$ of rank $\D$.   However, for $\D>8$, there are multiple isomorphism classes of such lattices, and all isomorphism classes 
can appear.   At the point in moduli space at which $\Lambda_+$ appears, the CFT partition function is
\be\label{wiffo}\frac{\Theta_{\Lambda_+}(\tau)\bar\Theta_{\Lambda_-}(\tau)}{|\eta(\tau)|^{2D}}. \ee
Here $\Theta_{\Lambda_+}$ and $\Theta_{\Lambda_-}$ are holomorphic theta functions associated to the lattices $\Lambda_+$ and $\Lambda_-$;
$\Theta_{\Lambda_-}$ is complex-conjugated because we have assumed the $\Lambda_-\subset V_-$ so that the $\Lambda_-$ modes are right-moving.

In this situation, the only possible averaging is over the choice of even integer unimodular lattice $\Lambda_+$; there are finitely many possibilities, depending on $\D$.\footnote{The number of such lattices is finite, but grows rapidly with $\D$.  For example, at $\D=48$ there are at least $10^{120}$ such lattices, although the number is not known exactly. At large $\D$, the number of even integer unimodular lattices  up to isomorphism grows like $\D^{\D^2}$.}
The holomorphic case of the Siegel-Weil formula says that the average\footnote{In this averaging, one weights the contribution of a given lattice $\Lambda_+$ by
the inverse of the order of its automorphism group.}    of $\Theta_{\Lambda_+}$ over all possibilities is a holomorphic Eisenstein series of weight $\D/2$:
\be\label{pilo} \E_{\D/2}(\tau)=\sum_{(c,d)=1}\frac{1}{(c\tau+d)^{\D/2} }. \ee
Thus the average partition function of this class of theories is 
\be\label{nilo}\langle Z_\Sigma(\tau)\rangle_{\Lambda_-} =\frac{\E_{\D/2}(\tau)\bar\Theta_{\Lambda_-}(\tau)}{|\eta(\tau)|^2}. \ee
The symbol $\langle~~\rangle_{\Lambda_-}$ represents an average under the constraint  $\Lambda\cap V_-\cong \Lambda_-$.

To interpret this result from a gravitational point of view, we start with a seed partition function on $D\times S^1$, which we take to be the partition function
of the extended chiral algebra.  In the present example, this  is $\bar\Theta_-(\tau)/|\eta(\tau)|^{2\D}$.  We write this as
\be\label{hoddo} \frac{1}{\eta(\tau)^{\D}} \frac{\bar\Theta_-(\tau)}{\bar\eta(\tau)^\D}.  \ee
To simplify the remaining derivation, let us assume that $\D$ is divisible by 24 and not just by 8.   Then $\frac{\bar\Theta_-(\tau)}{\bar\eta(\tau)^\D}$ is modular-invariant,
and so we just have to sum over the modular images of the function $1/\eta(\tau)^{\D}$.    For $\D$ a multiple of 24, the subtle
24$^{th}$ roots of unity that appear in the modular transformation of $\eta(\tau)$  disappear, and we have  just
\be\label{noddor} \eta((a\tau+b)/(c\tau+d))^\D=(c\tau+d)^{\D/2}\eta(\tau).\ee
With this, we see immediately that the sum over modular images of $1/\eta(\tau)^\D$ is $\E_{\D/2}(\tau)/\eta(\tau)^\D$.   So the sum over modular
images of the gravitational expression in eqn. (\ref{hoddo}) does give the formula (\ref{nilo}) for the average partition function.   If we had assumed that $\D$
is divisible by 8 but not necessarily by 24, we would have reached the same result after analyzing and  canceling some cube roots of unity.

In this discussion, we started with the Narain family of CFT's, based on an even integer unimodular lattice $\Lambda$.    Upon assuming an enhancement
of the chiral algebra, we restrict to a sublattice of $\Lambda$.  Such a sublattice automatically is still even and integer but possibly 
not unimodular.   So the averaging involves
the Siegel-Weil formula for a general even integer lattice.   Alternatively, to study a family of spin CFT's, which depend on a spin structure on $\Sigma$,
we could start with an integer lattice $\Lambda$ that is unimodular but not even.   After assuming an enhancement of the chiral algebra, $\Lambda$ would
be replaced by an integer sublattice that generically is neither even nor unimodular.    So the averaging in this case would depend on the Siegel-Weil formula
for a general integer lattice.

\section{Higher Genus And Disconnected Boundaries}\label{highg}

\subsection{Higher Genus}\label{hg}

We will now describe the Siegel-Weil formula at higher genus, and understand its interpretation in terms of our conjectured exotic theory of gravity.  The higher genus CFT partition function is more complicated, in part because a surface $\Sigma$ of genus $g>1$ does not admit
a flat metric, and hence in any explicit formula there is no way to avoid the conformal anomaly.    We will therefore need to be more schematic.

The genus $g$ partition function of a CFT in the Narain family can be written as 
\be\label{ombo} Z_\Sigma(m,\tau)= 
\frac{\Theta(m,\tau)}{\Phi},\ee
 where $\Theta(m,\tau)$ comes from a momentum-winding sum
and is the Siegel-Narain theta function in genus $g$, and $\Phi$ comes from the integral over oscillator modes.   
As before, $m$ denotes a point in the CFT moduli space $\M_\D$; $\tau$  now represents the whole set of moduli of $\Sigma$. 

Since the denominator $\Phi$ is not sensitive to the CFT moduli, averaging over $\M_\D$ means averaging 
$\Theta(m,\tau)$ over $\M_\D$.    This average is described again by a Siegel-Weil formula.   Using the higher genus analog of the Siegel-Weil formula,\footnote{In Appendix \ref{diffeq}, we discuss the derivation of the Siegel-Weil formula at genus $g$, by generalizing
 the method presented in the previous section.} the
result can be written as
\be\label{wilpo} \langle Z_\Sigma(m,\tau)\rangle = \frac{E_{\D/2}(\tau)}{(\det\,\Im\,\Omega)^{\D/2}|\det'\bar\partial|^{\D}},\ee
where $E_{\D/2}(\tau)$, to be described shortly, is an Eisenstein series of the group $\Sp(2g)$; $\Omega$ is the period matrix of $\Sigma$;
 and $\det'\bar\partial$ is the determinant of the
$\bar\partial$ operator of $\Sigma$, mapping functions to $(0,1)$-forms, with zero-modes removed.  In the denominator on the right hand side of eqn. (\ref{wilpo}), the
factor $(\det\,\Im\,\Omega)^{\D/2}$ generalizes $\tau_2^{\D/2}$ in eqn. (\ref{hubb}), and $|\det'\,\bar\partial|^{\D}$ generalizes $|\eta(\tau)|^{2\D}$ in that formula.

Our main interest here, however, is the Eisenstein series that appears in the numerator.   Let us first restate in an alternative way the definition
of the real analytic Eisenstein series that we used in genus 1.    If $\Sigma$ has genus 1, then the lattice $\Gamma\subset H_1(\Sigma,\Z)$ is a copy of $\Z^2$.   Any primitive
 rank
1 sublattice $\Gamma_0\subset \Gamma$ is a Lagrangian sublattice.\footnote{Lagrangian sublattices were defined in the introduction; for the definition 
of a primitive lattice, see footnote \ref{full}.}   Once we pick a basis of $\Gamma$, say by choosing an $A$-cycle $\A$
and a $B$-cycle $\B$ on $\Sigma$, $\Gamma_0$ can be specified by giving its generator, which is a linear combination
$c\A+d\B$,  with relatively prime integer coefficients $c,d$, up to sign.     Hence the sum over such pairs in the
definition (\ref{nifg}) of the Eisenstein series can be interpreted as a sum over Lagrangian sublattices.

The genus $g$ analog of the Siegel-Weil formula similarly involves a sum over Lagrangian sublattices.    Once we pick a Lagrangian sublattice $\Gamma_0\subset
\Gamma$, it is possible to define $\det\,\Im\,\Omega$ without any additional choices.   To do this, we first pick a set of $A$-cycles $\A^i$ that provide a basis
of  $\Gamma_0$,
and a complementary set of $B$-cycles $\B_j$, the nonzero intersection pairings being $\la \A^i,\B_j\ra=\delta^i_j$.    The homology classes of the $\B_i$
are not uniquely determined, but they are determined up to $\B_i\to \B_i+n_{ij}\A^j$, $n_{ij}\in\Z$.   Then one picks a basis of holomorphic 1-forms $\omega_k$
with $\oint_{\A^i}\omega_j=\delta^i_j$, and defines the period matrix by $\Omega_{ij}=\oint_{\B_i}\omega_j$.    A shift  $\B_i\to \B_i+n_{ij}\A^j$ shifts the period matrix
by $\Omega_{ij}\to\Omega_{ij}+n_{ij}$, without changing $\Im\,\Omega$.    Replacing the chosen $\A^i$ by a different basis of the same lattice $\Gamma_0$ changes
$\Omega$ to $P\Omega P^{\mathrm{tr}}$, where $P$ is an integer-valued matrix of determinant $\pm 1$ ($P^{\mathrm{tr}}$ is the transpose of $P$), without 
affecting $\det\,\Im\,\Omega$.    So in short $\det\,\Im\,\Omega$ is well-defined once $\Gamma_0$ is chosen.   

   For a given Lagrangian sublattice $\Gamma_0$, let $\det\,\Im\,\Omega_{\Gamma_0}$ be the
corresponding value of the determinant of the imaginary part of the period matrix.     Then the definition of the Eisenstein series is
\be\label{belfo} E_s(\tau) =\sum_{\Gamma_0} \left(\det\,\Im\,\Omega_{\Gamma_0}\right)^s.\ee
The sum runs over all Lagrangian sublattices.
For $g=1$, $\Omega$ is the $1\times 1$ matrix $\tau=\tau_1+\i\tau_2$, so $\det\,\Im\,\Omega=\Im\,\Omega=\tau_2$.    Hence (\ref{belfo}) reduces for $g=1$
to the sum over modular images of $\tau_2^s$.   This is the definition that we used in eqn. (\ref{nifg}), though in that case we wrote an explicit formula
for the dependence of $\Im\,\tau$ on the choice of Lagrangian sublattice.   
 It is possible to do the same for any $g$, and rewrite eqn. (\ref{belfo}) as a sum over modular images just as in eqns. (\ref{nifg}) and (\ref{eismod}).
This version of eqn (\ref{belfo}), where the Eisenstein series is written explicitly as a sum over $\Sp(2g,\Z)$, is given in equation (\ref{esdef}).

Now we can explain the properties that a hypothetical bulk dual of the average over Narain moduli space should have in order to reproduce the result
(\ref{wilpo}) for the average partition function.    The bulk contributions to the path integral should be labeled by Lagrangian sublattices $\Gamma_0$.
In terms of classical geometry, we might try to attribute these contributions to handlebodies with boundary $\Sigma$, since (for connected $\Sigma$)  
these are in natural correspondence with Lagrangian sublattices, as noted in the introduction.   That viewpoint will not work well in the disconnected case,
which we come to in section \ref{disconnected}, so instead we will just say that the bulk contributions are labeled by Lagrangian sublattices.
The bulk path integral for a given $\Gamma_0$ should be
\be\label{poggo} \frac{1}{|\det_{\Gamma_0}'\,\bar\partial|^{\D}} ,\ee
where we note that the determinant depends on $\Gamma_0$ because of the subtleties involved in treating the kernel and cokernel of $\bar\partial$.   
Writing this as
\be\label{novo} \frac{1}{(\det\,\Im\,\Omega_{\Gamma_0})^{\D/2}|\det_{\Gamma_0}'\,\bar\partial|^{\D}}\cdot (\det\,\Im\,\Omega_{\Gamma_0})^{\D/2},\ee
where the denominator  $(\det\,\Im\,\Omega_{\Gamma_0})^{\D/2}|\det_{\Gamma_0}'\,\bar\partial|^{\D}$   actually
does not depend\footnote{The expression $1/\left((\det\,\Im\,\Omega_{\Gamma_0})|\det'_{\Gamma_0}\,\bar\partial|^2\right)^{\D/2}$ is actually the partition function,
per unit volume in the target space, of a sigma-model with target $\R^D$.  Thus in particular it does not depend on the choice of $\Gamma_0$.
 See for example \cite{M,N}.  We will not explore this rather subtle point here as our
interest in the present paper is really in the numerator of the partition function.} on $\Gamma_0$, we see that to get the full partition function, we just need to sum $ (\det\,\Im\,\Omega_{\Gamma_0})^{\D/2}$ over
the choice of $\Gamma_0$.    But this  sum is the definition of the Eisenstein series $E_{\D/2}(\tau)$, so if (\ref{poggo}) is the appropriate formula for the contribution
of a given $\Gamma_0$ to the path integral, then the sum over all $\Gamma_0$ will indeed reproduce the desired answer (\ref{wilpo}) for the average partition
function.   

Eqn. (\ref{poggo}) is a fairly plausible formula for the handlebody path integral in a theory in which
the only physical degrees of freedom are the boundary current algebra modes, the analogs of the Brown-Henneaux modes for gravity.   
Such boundary current algebra  modes
correspond to $D$ left- and right-moving massless scalars that lack zero-modes, and eqn. (\ref{poggo}) is a natural candidate for the path integral for such
fields.  One can think of $ \frac{1}{|\det_{\Gamma_0}'\,\bar\partial|^{\D}} $ as a particular conformal block for the $\U(1)^{2\D}$ current algebra.
This conformal block can be characterized by saying that what is propagating through any one-cycle that represents a class in $\Gamma_0$ is the vacuum
module of the current algebra.
In section \ref{gaugetheory}, we do a direct gauge theory calculation that, under certain assumptions,
exhibits $ \frac{1}{|\det_{\Gamma_0}'\,\bar\partial|^{\D}} $ as a gauge theory
partition function in the handlebody.

 The justification to consider only handlebodies is thin, as acknowledged in the introduction, unless we assume that we are studying an exotic theory
of gravity in which ``manifolds with boundary $\Sigma$'' are classified entirely by the associated Lagrangian sublattice $\Gamma_0\subset H_1(\Sigma,\Z)$.   

One more remark may provide some background for our discussion of the case that the boundary is not connected.    The hypothetical theory that we are discussing
is not conventional gravity and does not have a conventional semi-classical limit.    The closest analog is to consider $\D$ to be large.  
For generic $\tau$  there will be one Lagrangian sublattice $\Gamma_0$ that maximizes $\det\,\Im\,\Omega_{\Gamma_0}$.   For large $\D$, this particular
Lagrangian sublattice then makes the dominant contribution in the definition (\ref{belfo}) of the Eisenstein series $E_{\D/2}$.    Other contributions are
exponentially suppressed.  Of course, as we vary $\tau$, there will be large $\D$ phase transitions at which two Lagrangian sublattices exchange dominance.

 We will mention one additional subtlety which appears when we work at finite $\D$ rather than in the large $\D$ limit. 
This can be seen by investigating the Eisenstein series $E_{\D/2}(\tau)$ which appears as the average of the Siegel-Narain theta function.  At genus 1, we saw that the Eisenstein series diverged unless we took $D>2$; this reflected the fact that the integral over $\M_\D$ of the Siegel-Narain theta function was divergent.  It turns out that the genus $g$ version of the Eisenstein series diverges unless
\be D<g+1. \ee
This and other properties of $E_{D/2}(\tau)$ are discussed in more detail in Appendix \ref{diffeq}.  As in the genus one case, this reflects a genuine divergence of the averaging over Narain moduli space.
At finite $\D$, the average partition function diverges for sufficiently large $g$: when this happens, the typical CFT lives ``at the boundary" of $\M_\D$.  The result is that the hypothetical gravitational dual theory  can compute relatively coarse averaged CFT observables -- namely, the low genus partition functions which encode the average spectrum and low moments of the OPE coefficients -- but fails to compute highly refined observables, such as the large $g$ partition functions which compute higher moments of the OPE coefficients.  

This has interesting implications for the structure of non-perturbative effects in our theory of gravity.  As noted earlier (and described in more detail in section \ref{gaugetheory}), our bulk Chern-Simons theory is one-loop exact at all genus, and so accounts for all of the perturbative effects which arise in a large $\D$ limit.  The Eisenstein series then computes a set of non-perturbative corrections.  However, we see that at finite $\D$ this is still not enough, as this sum diverges at sufficiently large genus.  This may hint that further non-perturbative effects are necessary in order to render the theory sensible at finite $\D$, analogous to the ``doubly non-perturbative effects" \cite{SSS} which are necessary in JT gravity in order to render the theory sensible nonperturbatively.\footnote{We note that, although they diverge, the relevant Eisenstein series can be formally defined by analytic continuation for $g>D+1$. This may aid in interpreting our results at finite $\D$.}

\subsection{Disconnected Boundaries}\label{disconnected}

As explained in the introduction, one is particularly interested to know what is the outcome of the averaging procedure if $\Sigma$ is not connected.

Recall first that the period matrix $\Omega$ of a genus $g$ Riemann surface $\Sigma$ is a $g\times g$ symmetric complex-valued matrix whose imaginary
part is positive-definite. In what follows, $\Omega$ always refers to a complex matrix with those properties.   For genus 
$g>3$, it is not true that any such $\Omega$ is the period matrix of some $\Sigma$.   In general, such an $\Omega$
is associated to a principally polarized abelian variety of rank $g$ which is not necessarily the Jacobian of any $\Sigma$.

However, all the formulas of section \ref{hg} make sense for an arbitrary $\Omega$, whether or not it is the period matrix of a Riemann surface.
For example, the Siegel-Narain theta function $\Theta(m,\tau)$ is defined by a momentum-winding sum that depends on $\Sigma$ only through its period
matrix $\Omega$.     The only properties of $\Omega$ that are needed for this sum to make sense are that it is symmetric and has positive-definite imaginary part.\footnote{Properties of the space of such matrices, as well as an explicit formula for the Siegel-Narain theta function, are given in Appendix \ref{diffeq}.}
(Positivity is needed for convergence of the momentum-winding sum.)    To emphasize this, we could denote the theta function as $\Theta(m,\Omega)$ rather than
$\Theta(m,\tau)$.    Moreover, the Siegel-Weil formula for averaging over $m$ holds for an arbitrary $\Omega$, not necessarily the period matrix of any Riemann
surface.   In fact, in the mathematical literature it is not usual to restrict $\Omega$ to be a period matrix.

This being so, before discussing disconnnected surfaces, we might want to ask if we can find a physical interpretation of the Siegel-Weil formula for an $\Omega$
that is not associated to a Riemann surface.  Can we generalize the question that we have been asking so that the answer will involve the more general case of the
Siegel-Weil formula?   We can, though this involves asking a question that is possibly less natural than the question that we have been asking so far about the
average of the partition function.    Let $C_\alpha$, $\alpha=1,\cdots,2g$ be loops in $\Sigma$ that represent a basis of $H_1(\Sigma,\Z)$, and let $X^p$, $p=1,\cdots,\D$
be the scalar fields of a Narain model.    We add to the action a bilocal term
\be\label{wilop} \sum_{\alpha\beta pq} d_{\alpha\beta pq}\oint_{C_\alpha}\d X^p \oint_{C_\beta}\d X^q ,\ee with arbitrary coefficients $d_{\alpha\beta pq}$.
This has no effect on the set of classical solutions of the theory, and no effect on the the quantum oscillations around a classical solution.   But it changes
the action of a classical solution.   By suitably adjusting the coefficients, we can arrange so that the momentum-winding sum is $\Theta(m;\Omega)$ for any
desired $\Omega$.   So averaging over $m$ in this situation will involve the Siegel-Weil formula for arbitrary $\Omega$.   The reader may or may not consider
this a compelling context for the more general Siegel-Weil formula.

Regardless, a special case of the fact that the Siegel-Weil formula holds for any $\Omega$ is that it holds for any $\Sigma$, connected or not.
For example, suppose that $\Sigma$ is the disjoint union of two connected surfaces $\Sigma'$ and $\Sigma''$, of genus $g'$ and $g''$, and whose moduli we denote as $\tau'$ and $\tau''$.   Set ${ g}=g+g'$ and write $\tau$ for the whole collection of moduli $\tau,\tau'$.
      For fixed $m\in\M_\D$, the partition function on $\Sigma$ is a product:
\be\label{olko}Z_\Sigma(m,\tau)=Z_{\Sigma'}(m,\tau') Z_{\Sigma''}(m,\tau'').\ee
We want to average over $m$ and compute the connected correlation function $\langle   Z_{\Sigma'}(m,\tau') Z_{\Sigma''}(m,\tau'')\rangle_c$.
The function that we need to average is, from eqn. (\ref{ombo}),
\be\label{nolko} \frac{\Theta(m,\tau')\Theta(m,\tau'')}{\Phi_{\Sigma'}\Phi_{\Sigma''}}.\ee
It is the numerator that has to be averaged, since only the numerator depends on $m$.

Let  $\Gamma'=H_1(\Sigma',\Z)$, $\Gamma''=H_1(\Sigma'',\Z)$, and $\Gamma=\Gamma'\oplus \Gamma''
=H_1(\Sigma,\Z)$.   On $\Gamma$, there is an intersection pairing, which is simply the sum of the intersection pairings on $\Gamma'$ and on $\Gamma''$.
A Lagrangian sublattice $\Gamma_0$ of $\Gamma$ is  a\footnote{Primitive, as in footnote  \ref{full} in the introduction.}
rank $g$ sublattice of $\Gamma$ on which the intersection pairing vanishes.   Such a sublattice
 may be the direct sum of Lagrangian sublattices $\Gamma_{0}'\subset \Gamma'$ and $\Gamma_{0}''
\subset \Gamma''$, in which case we will say that $\Gamma_0$ is decomposable.  But this is not the only possibility.   
There is no problem to define a period matrix of $\Sigma$ associated to a Lagrangian sublattice
that is not decomposable.   We will work out an example shortly.   

Let $\Omega'$ and $\Omega''$ be the period matrices of $\Sigma'$ and $\Sigma''$.    Then the direct sum\footnote{Let $\h\Sigma$ be a connected
Riemann surface defined as the connected sum of $\Sigma'$ and $\Sigma''$.   In a limit that $\h\Sigma$ degenerates to the union of $\Sigma'$ and $\Sigma''$
joined at a point, the period matrix of $\h\Sigma$ reduces to that of the disconnnected surface $\Sigma$ (eqn. (\ref{jud})).   This fact actually gives one way to prove that the Siegel-Weil formula must
apply to disconnected Riemann surfaces if it applies to connected ones.    But this is not very helpful in understanding the geometric meaning of the averaged path
integral on a disconnected manifold, because a generic handlebody with boundary $\h\Sigma$ is not related in a nice way to a three-manifold whose boundary
is the disjoint union of $\Sigma'$ and $\Sigma''$.   The problem arises precisely in the interesting case of indecomposable Lagrangian sublattices (see below).}
\be\label{jud} \Omega =\begin{pmatrix}\Omega'& 0 \cr 0 & \Omega''\end{pmatrix}, \ee
which we define as the period matrix of $\Sigma$, is symmetric with positive-definite imaginary part, 
 so we can apply the Siegel-Weil formula to the corresponding Siegel-Narain theta function.    But this
theta function is just a product: 
\be\label{nud} \Theta(m,\Omega)=\Theta(m,\Omega')\Theta(m,\Omega'')\ee
because the momentum-winding sum of a disjoint union of Riemann surfaces is just the product of the two separate momentum-winding sums.  
The right hand side is the function that we want to average in order to compute $\la Z_{\Sigma'}(m,\tau') Z_{\Sigma''}(m,\tau'')\ra$, and the left hand side
is the function that we know how to average using the Siegel-Weil formula.   Applying the Siegel-Weil formula, we learn that 
\be\label{rud} \langle Z_{\Sigma'}(m,\tau') Z_{\Sigma''}(m,\tau'')\rangle =   \frac{E_{D/2}(\tau',\tau'')}
{\left((\det\,\Im\,\Omega')^{\D/2}|\det'\bar\partial_{\Sigma'}|^\D\right)   \left((\det\,\Im\,\Omega'')^{\D/2}|\det'\bar\partial_{\Sigma''}|^{\D}\right)    } . \ee
The denominator in this formula is the product of the denominators in the usual expressions for $Z_{\Sigma'}$ and $Z_{\Sigma''}$;  as usual it
 only depends on the moduli $\tau',\tau''$, and not on the
choices that are used to define the period matrices and determinants.   
The definition of the Eisenstein series is as usual
\be\label{lud} E_{\D/2}(\tau',\tau'')=\sum_{\Gamma_0} \left(\det \Im \,\Omega_{\Gamma_0}\right)^{\D/2} .\ee
The sum runs over all Lagrangian sublattices $\Gamma_0\subset \Gamma$, and $\Omega_{\Gamma_0}$ is the period matrix defined using $\Gamma_0$.   
If we restrict the sum to the decomposable case  $\Gamma_0=\Gamma_{0}'\oplus \Gamma_{0}''$,
where the summands are Lagrangian sublattices of $\Gamma'$ and $\Gamma''$, respectively, then the right hand side will reduce to
$E_{\D/2}(\tau')E_{\D/2}(\tau'')$.    When inserted in eqn. (\ref{rud}), this will give the disconnected contribution to the correlation function.
The connected correlator comes precisely from Lagrangian sublattices that are not decomposable.

To make this more concrete, we will describe an explicit example of an indecomposable  Lagrangian sublattice $\Gamma_0$ 
and  compute its contribution to the connected
correlator.   Let $\Sigma'$ and $\Sigma''$ be Riemann
surfaces of genus 1, with respective modular parameters $\tau'$ and $\tau''$.   On $\Sigma'$, we pick an $A$-cycle $\a'$ and a $B$-cycle $\b'$; on
$\Sigma''$ we pick an $A$-cycle $\a''$ and a $B$-cycle $\b''$.   The nonzero intersection numbers are
\be\label{hbo}\la \a',\b'\ra=\la \a'',\b''\ra=1. \ee
We also pick holomorphic differentials $\omega'$ on $\Sigma'$ and $\omega''$ on $\Sigma''$ normalized so that
\be\label{nbo}\oint_{\a'}\omega'=\oint_{\a''}\omega''=1,~~~\oint_{\b'}\omega'=\tau',~~~~\oint_{\b''}\omega''=\tau''. \ee
We now want to pick a Lagrangian sublattice $\Gamma_0$.    A sublattice generated by, for example, $\a'$ or $\b'$ along with $\a''$ or $\b''$ is decomposable.   Instead
we pick one generated by
\be\label{zbo}\A^1=\a'-\a'',~~~~ \A^2=\b'+\b''. \ee
The minus sign in the definition of $\A^1$ ensures that $\la \A^1,\A^2\ra=0$, so that $\A^1$ and $\A^2$ indeed generate a Lagrangian sublattice.
For a dual pair of cycles, we can pick
\be\label{zobo}\B_1=\b',~~~~ \B_2=-\a''. \ee
This ensures that $\la\B_1,\B_2\ra=0$ and
\be\label{kobo} \la \A^i,\B_j\ra=\delta^i_j. \ee

To compute the period matrix, we need  holomorphic differentials $\omega_i$ with
\be\label{lobo}\oint_{\A^i}\omega_j=\delta^i_j. \ee
These are
\be\label{hobo}\omega_1=\frac{\tau''\omega'-\tau'\omega''}{\tau'+\tau''},~~~\omega_2=\frac{\omega'+\omega''}{\tau'+\tau''}.\ee
The period matrix will then be
\be\label{gobo}\Omega_{ij}=\oint_{\B_i}\omega_j. \ee
So
\begin{align}\label{permat} \Omega_{11} &= \oint_{\B_1}\omega_1= \frac{\tau'\tau''}{\tau'+\tau''}\cr
   \Omega_{12}&=\oint_{\B_1}\omega_2=\oint_{\B_2}\omega_1 =  \frac{\tau'}{\tau'+\tau''}\cr
   \Omega_{22}&=\oint_{\B_2}\omega_2 =-\frac{1}{\tau'+\tau''} .\end{align}
   
Expanding in real and imaginary parts by  $\tau'=\tau_1'+\i\tau_2'$, $\tau''=\tau_1''+\i\tau_2''$, we find that
\be\label{jojobo}\Im\,\Omega=\frac{\tau_2'}{|\tau'+\tau''|^2}\begin{pmatrix} |\tau''|^2 & \tau_1''\cr\tau_1''&1\end{pmatrix} +\frac{\tau_2''}{|\tau'+\tau''|^2}
\begin{pmatrix} |\tau'|^2 &-\tau_1' \cr -\tau_1' & 1\end{pmatrix},\ee 
which is positive-definite, as expected.   There is a simple result for $\det\,\Im\,\Omega$:
\be\label{polobo}\det\,\Im\,\Omega=\frac{\tau_2'\tau_2''}{(\tau_1'+\tau_1'')^2+(\tau_2'+\tau_2'')^2}.\ee
The contribution of this particular Lagrangian sublattice to the Eisenstein series is
\be\label{wolobo}\left(\det\,\Im\,\Omega\right)^{\D/2}=\left(\frac{\tau_2'\tau_2''}{(\tau_1'+\tau_1'')^2+(\tau_2'+\tau_2'')^2}\right)^{\D/2}.\ee

Intuitively, one expects that in anything that one would call a semiclassical limit,
connected correlators between different components of $\Sigma$ should be small.  As remarked near the end of section \ref{hg}, in the present context,
the closest analog of a semiclassical limit is large $\D$.   In fact, the connected contribution that we have analyzed is exponentially small for large
$\D$ compared to the disconnected correlation function.   To see this, note that eqn. (\ref{polobo}) implies a general upper bound
\be\label{oloboi}\det\,\Im\,\Omega\leq \frac{1}{4},\ee
where the maximum is attained if and only if $\tau_1'+\tau_1''=0$, $\tau_2'=\tau_2''$.   Therefore, the contribution to the Eisenstein series
from this particular indecomposable sublattice  is at most
$ \left(\frac{1}{4}\right)^{\D/2}$.
However, there is always a decomposable Lagrangian sublattice whose contribution to the Eisenstein series 
is at least $\left(\frac{3}{4}\right)^{\D/2}$.   Indeed, the contribution of
the decomposable sublattice generated by $\A'$ and $\A''$ is $\tau'_2{}^{\D/2}\tau''_2{}^{\D/2}$.   If $\tau',\tau''$ are in the usual fundamental domain for $\SL(2,\Z)$,
then $\tau_2',\tau_2''\geq \sqrt 3/2$, and the corresponding contribution to the Eisenstein series is 
$\tau'_2{}^{\D/2}\tau''_2{}^{\D/2}\geq \left(\frac{3}{4}\right)^{\D/2}$.   Even if
$\tau'$ and $\tau''$ are not in the usual fundamental domain, by acting on $\A'$ and on $\A''$ with separate $\SL(2,\Z)$ transformations that map
$\tau'$ and $\tau''$ into the usual fundamental domain, we find a different decomposable Lagrangian sublattice whose contribution is at least 
$\left(\frac{3}{4}\right)^{\D/2}$.
So the connected contribution to the correlator that we have examined is smaller than the disconnected correlator by at least a factor of
$3^{\D/2}$.   One expects that all connected contributions are similarly exponentially suppressed for large  $\D$.    The interpretation in terms of a hypothetical
bulk dual theory would be that ``manifolds''    with ``wormhole'' connections between different boundary components have larger action  (or at least smaller
quantum path integrals) than disconnected ``manifolds.''

The connected correlator that we have analyzed has the interpretation of \be\label{concor}\left\langle \Tr\,\exp(\i \tau_1' P-\tau_2' H) \,\Tr\,\exp(\i\tau_1''P-\tau_2''H)\right\rangle,\ee where the traces are taken in the CFT Hilbert space, 
and $H$ and $P$ are the CFT Hamiltonian and momentum operators.   This correlator is a real-analytic function of $\tau_1',\tau_2',\tau_1''$, and $\tau_2''$.
so it can be analytically continued to complex values of those variables, at least within certain limits.   In particular, to get an analog of  the ``spectral form factor,''
we can set $\tau_2'=\beta+\i t$, $\tau_2''=\beta-\i t$, where $\beta$ and $t$ are both real; $t$ is interpreted as a real time parameter.   The limit of large $t$, keeping fixed $\beta$, $\tau_1'$, $\tau_1''$
probes interesting properties of the spectrum, and has been investigated in detail in other models; for example see \cite{CC}.     It is not difficult to
calculate the contribution of the indecomposable sublattice $\Gamma_0$ to the spectral form factor.    For $\Sigma'$, $\Sigma''$ both of genus 1, the denominator in the
formula (\ref{rud}) for the correlation function simplifies to $(\Im\,\tau'\,\Im\,\tau'')^{\D/2}|\eta(\tau')\eta(\tau'')|^\D$.    Since  $\eta(\tau)$ is holomorphic
and  $|\eta(\tau+1)|=|\eta(\tau)|$, it follows that, when we give $\tau'_2$, $\tau''_2$ imaginary parts $\pm \i t$,
 $|\eta(\tau')\eta(\tau'')|$ is periodic in $t$ with period 1.   This  periodicity simply reflects the fact that the energy differences between current algebra modes
are integer multiples of $2\pi$. 
  The current algebra modes are unaffected by  averaging over Narain moduli space, so after this averaging
the correlation function retains the periodic factor $1/|\eta(\tau')\eta(\tau'')|^\D$.   More interesting is the $t$ dependence of the averaged product of Narain theta
functions.   This is
\be\label{wolo}  \frac{1}{(\tau_2'\tau_2'')^{\D/2}}\sum_{\Gamma_0} (\det\,\Im \,\Omega_{\Gamma_0})^{\D/2}. \ee
In view of eqn. (\ref{wolobo}), the contribution to this expression of the particular indecomposable Lagrangian sublattice that we considered is actually a positive constant independent
of $t$.    %Since all contributions are positive, this is enough to show that the spectral form factor does not vanish for $t\to\infty$.  We expect that it has a constant limit, like the particular contribution that we evaluated.   
We expect that the full spectral form factor has a positive constant
limit at $t\to\infty$, just like the particular contribution that we have evaluated.
We note that, as emphasized in  \cite{CC}, for the spectral form factor to approach a non-zero constant at late time is a key signature of the discreteness of the spectrum; this discreteness is, in general, quite difficult to see in a quantum gravity computation.

One may ask for a geometric realization of the Lagrangian sublattice that we have considered.    It is actually not difficult to find one.   
We want an oriented three-manifold $Y$ whose boundary consists of the disjoint union of $\Sigma'$ and $\Sigma''$, such that $\A'-\A''$ and $\B'+\B''$ 
are boundaries in $Y$.   We can take $Y=S^1\times S^1\times I$, where $I$ is a unit interval. The two ends of $I$ correspond to the two boundaries of $Y$,
each of which is a copy of $S^1\times S^1$.   We identify $\Sigma'$ and $\Sigma''$ with the two boundaries
of $Y$ in such a way that $\A'$ and $\A''$ are identified with the first factor of $S^1\times S^1$ (in the first and second boundary of $Y$, respectively)
and $\B'$ and $-\B''$ are similarly identified with the second factor.
The reason for a minus sign in the statement about $\B''$ is that $Y$ has to be oriented and its orientation has to induce the orientations
of the boundaries $\Sigma'$ and $\Sigma''$ that were built into the statement $\la\A',\B'\ra=\la\A'',\B''\ra=1$. (The need for compatible orientations
was noted following eqn. (\ref{pilk}).)    
So the identification  of $\Sigma'$ and $\Sigma''$ with the
boundaries of $Y$ has to involve a relative orientation reversal.  We used an orientation-reversing map of $\Sigma''$ that maps $(\A'',\B'')\to (\A'',-\B'')$.
We have given the simplest example of a $Y$ that is associated to the Lagrangian sublattice that we considered, but there are infinitely many others.  
Since there appears to be no natural way to get the answer $(\det\,\Im\,\Omega)^{\D/2}$ from a sum over distinct $Y$'s, it was suggested in the introduction that in the exotic
theory of gravity that is dual to an average over Narain moduli space, there is not a well-defined distinction between different $Y$'s that are associated to the
same Lagrangian sublattice.

We conclude by noting that, as with the connected case described in the previous subsection, the Eisenstein series which computes the analog of the sum over geometries in our theory of gravity does not necessarily converge when $\D$ is finite.  For example, we could consider the $N^{th}$ moment of the torus partition function:
\be
\langle Z_{\Sigma_1}(m,\tau_1)\dots Z_{\Sigma_N}(m, \tau_N) \rangle
\ee
where the $\Sigma_i$ are tori.  The Eisenstein series converges only when $N < \D-1$.  Thus at finite $\D$ our theory can successfully compute relatively coarse features of the spectrum of the CFT, namely those features encoded in low moments of $Z_\Sigma(m,\tau)$, but fails when $N$ is large compared to $D$.  This may indicate the need to include further non-perturbative effects.

Perhaps some intuition may be gained in this case by thinking of $Z$ not as a random CFT partition function, but rather as a random $\D \times \D$ matrix (a reasonable analogy, since $Z$ comes from a random rank $\D$ lattice).  Only the first $D$ moments of such a matrix are independent of one another, and these first $D$ moments are the only data required to completely characterize the probability distribution.  It may be that the gravitational theory can only be used to compute the ``independent" pieces of data needed to characterize the distribution on the space of CFTs.  Having specified these data, the higher order observables are then completely determined.  This is reminiscent of the notion of gravitational null states appearing in \cite{Marolf:2020xie} (see also \cite{Maloney:2015ina}).

\subsection{Genus Zero}\label{genuszero}

We started this paper in genus 1, skipping the basic case of a surface of genus 0.    Here we will make amends for this omission.

If $\Sigma_0$ is a surface of genus 0, then $\Sigma_0$ has no complex moduli.   Moreover, 
the partition function of the Narain CFT on $\Sigma_0$ is  $Z_{\Sigma_0}(m)=1$, independent of $m$.  
So averaging $Z_{\Sigma_0}(m)$ over $m$ will not have any effect.

From the point of view of the Siegel-Weil formula, since $H_1(\Sigma_0,\Z)=0$, there is no nontrivial sum to be carried out in averaging $Z_{\Sigma_0}(m)$.
The only Lagrangian sublattice of the zero lattice $H_1(\Sigma_0,\Z)$ is the zero lattice.   Since  ``manifolds with boundary $\Sigma$'' in the theory under
discussion correspond  in general to Lagrangian sublattices of $H_1(\Sigma,\Z)$, we conclude that in the case of a surface of genus 0, 
there is only one ``manifold with boundary $\Sigma_0$.''     The closest analog of this object in classical geometry would be a three-ball, but in classical geometry
it is far from unique as an oriented manifold whose boundary is a surface of genus 0.

Since $Z_{\Sigma_0}(m)$ is a constant, independent of $m$, there is no connected correlator between $Z_{\Sigma_0}(m)$ and $Z_{\Sigma'}(m,\tau)$ for
any other surface $\Sigma'$.    In terms of the Siegel-Weil formula, one would interpret this as follows.  Let $\Sigma$ be the disjoint union of $\Sigma_0$ and $\Sigma'$.
Since $H_1(\Sigma,\Z)\cong H_1(\Sigma',\Z)$, ``manifolds with boundary $\Sigma$'' are in 1-1 correspondence with ``manifolds with boundary $\Sigma'$.''
It appears that if there really is an exotic gravitational theory with the properties suggested by the Siegel-Weil formula, then in this theory there is no
``wormhole'' connecting a genus 0 surface to anything else.   Alternatively, it may be that in this theory of gravity, there is no notion of whether spacetime
is connected and thus no way to say whether or not $\Sigma_0$ and $\Sigma'$ are connected through the wormhole.

\section{Path Integrals In Gauge Theory}\label{gaugetheory}

\subsection{Preliminaries}

In this section, we will compare the formulas of sections \ref{genusone} and \ref{highg} to direct evaluation of a gauge theory partition function
on a handlebody.   This will be done by adapting formulas in \cite{GMY}, where a similar direct calculation was done for Einstein gravity.    

Let us first recall that for $\U(1)$ gauge theory with the standard Maxwell action, the partition function is
\be\label{juggu}Z_{\mathrm{Max}} = \frac{\det'\,\Delta_0}{(\det'\,\Delta_1)^{1/2}}. \ee
Here $\Delta_0$ is the Laplacian acting on a field of spin 0, and $\Delta_1$ is the Laplacian acting on a vector field or 1-form.   With a standard gauge-fixing,
the denominator comes from the path integral over the gauge field, and the numerator is the ghost determinant.

Now consider a gauge field with Chern-Simons action.   First consider the case that the gauge group is $\U(1)$, with a single gauge field $A$,
 and the action on a three-manifold $Y$ is
\be\label{llugu}\frac{1}{2\pi} \int_Y A\wedge \d A. \ee
In the approach to quantization followed in \cite{Witten}, the gauge-fixing action is
\be\label{rugu}\frac{1}{2\pi}\int_Y \d^3x \sqrt g\left(\phi D_i A^i + \bar c D_i D^i c\right), \ee
where $c$ and $\bar c$ are ghost and antighost fields and $\phi$ is a scalar field that is a BRST partner of $\bar c$.    The
path integral for $c$ and $\bar c$ gives the usual ghost determinant $\det' \,\Delta_0$.  The kinetic operator acting on the bosonic fields $A,\phi$
can be regarded as the operator $L_-=*\d +\d *$ acting on differential forms of odd degree.   The corresponding path integral is $1/\sqrt{ \det' \,L_-}.$
But since $L_-^2$ is equivalent to the direct sum $\Delta_0\oplus \Delta_1$, we have $\det' \,L_-=(\det'\Delta_0\cdot \det'\Delta_1)^{1/2}$.  So finally
the relevant ratio of determinants for $\U(1)$ Chern-Simons theory is $(\det'\,\Delta_0)^{3/4}/(\det'\,\Delta_1)^{1/4}$.

The path integral of $\U(1)$ Chern-Simons theory is not just a product of determinants, as there is also a phase that involves an Atiyah-Patodi-Singer $\eta$-invariant
\cite{Witten}.   However, in the present paper we are interested in a $\U(1)^{2\D}$ theory based on a lattice $\Lambda$ of signature $(\D,\D)$, and in this
case, the phase cancels between modes on which the metric of $\Lambda$ is positive and modes on which it is negative. 
An even unimodular integer lattice $\Lambda$ of signature $(\D,\D)$ is actually simply the direct sum of $\D$ copies of a rank 2 lattice with intersection form
\be\label{intform} H=\begin{pmatrix} 0 & 1\cr 1 & 0 \end{pmatrix} . \ee
So the $\U(1)^{2\D}$ Chern-Simons theory based on $\Lambda$ is simply the product of $\D$ decoupled copies of a $\U(1)^2$ theory with two gauge fields $A,B$
and action
\be\label{wform}\frac{1}{2\pi}\int_Y A\wedge \d B. \ee
To the extent that the partition function of this theory can be calculated just by evaluating determinants, the determinants involved are the same ones as in the last
paragraph but with double the multiplicity, giving
\be\label{nform} Z_{\CS}=\frac{(\det'\,\Delta_0)^{3/2}}{(\det'\,\Delta_1)^{1/2}}. \ee
This formula was first obtained by A. Schwarz \cite{Schwarz}, who also recognized that this particular ratio of determinants is the Reidemeister-Ray-Singer torsion
(of a background flat connection, a trivial one in the case of expanding around $A=B=0$).  

Eqn. (\ref{nform}) is not the whole story for evaluation of the partition function of Chern-Simons theory.   It is not correct to simply ignore the zero-modes.
Zero-modes of $\Delta_1$ means that the classical solution about which we are expanding is part of a family, 
and zero-modes of $\Delta_0$ mean that
the gauge group has an unbroken subgroup of positive dimension.   In a full evaluation of the Chern-Simons path integral, we have to integrate over
the space of all classical solutions (summing in general over its connected components) 
 and divide by the volume of the unbroken gauge group.    Moreover, we are interested in evaluating the path integral for the case that $Y$ is a hyperbolic
 three-manifold with non-empty conformal boundary $\Sigma$.   In Chern-Simons theory, one usually requires a subtle analysis of the asymptotic behavior
 of the fields near the boundary, while in scalar field theory or Maxwell theory, one can just assume that perturbations vanish at infinity.
 
 Nevertheless, we will simply evaluate the right hand side of eqn. (\ref{nform}), for the case that $Y$ is a hyperbolic three-manifold with conformal boundary $\Sigma$.
 We do this using the formulas that were obtained in \cite{GMY} as part of a similar
 calculation for Einstein gravity.   These are formulas for the determinants of $\Delta_0$ and $\Delta_1$ in a space of perturbations that vanish at infinity.
We will find that this procedure works in the sense
that -- at least if $\Sigma$ is connected and $Y$ is a handlebody -- it gives the result that was needed in eqn. (\ref{poggo}) to provide a bulk dual to the
average over Narain moduli space.

Ideally, one would like to do a more rigorous evaluation of the Chern-Simons path integral and compare it precisely to the average
over Narain moduli space.   There seem to be real obstacles to this, as we discuss in section \ref{comparison}.   Ultimately, we do not
know to what extent gauge theory can be used to construct a bulk theory that is dual to an average over Narain moduli space.

\subsection{Path Integral On A Handlebody}\label{handleb}

We will evaluate determinants by a  heat kernel method, as in \cite{GMY}.   
The basic idea is that the determinant of an operator $\Delta$ is given by the following formula:
\begin{equation}
-\log \left(\det \Delta\right) = \int_{0^+}^\infty {dt \over t} {\rm Tr}\left(K_t\right)
\end{equation}
where $K_t\equiv e^{-t \Delta}$. % is the heat kernel associated with the operator $\Delta$.   
In the case of interest, where $\Delta$ is a differential operator on a manifold
$Y$, the trace is just an integral over $Y$, and $K_t$ can be found by solving the differential equation
\begin{equation}\label{heateq}
(\partial_t + \Delta)K_t=0,
\end{equation}  which we will momentarily write in position space as an equation for the heat kernel.
The  advantage of this technique   is that (\ref{heateq}) is a linear differential equation to which one can apply the method of images.  Thus by starting with $K_t$ on hyperbolic three-space ${\mathbb H}^3$, one can easily obtain $K_t$ on a general quotient ${\mathbb H}_3/\G$, where $\G$ is a discrete group
of automorphisms of $\HH^3$.    For simplicity, we will assume that every element of $\G$ other than the identity is of infinite order; this is so in many
interesting examples, including the groups (called Schottky groups) which are such that $\HH/\G$ is a handlebody.
 
{\blue As a simple illustration of this technique, let us  take our operator to be %consider the heat kernel for 
the scalar Laplacian $\Delta_0 \equiv -\nabla^2$ on ${\mathbb H}_3$, acting on the space of functions.
We then introduce the heat kernel $K^{{\mathbb H}_3}_t(x,x')$, which solves the equation}
\begin{equation}
(\nabla^2_x- \partial_t) K^{{\mathbb H}_3}_t(x,x')=0
\end{equation}
with the initial condition $K_t(x,x') = \delta(x,x')$  at $t=0$.  The solution is
\begin{equation}
K^{\mathbb H_3}_t(x,x') = {e^{-t-{d^2\over 4t}} \over (4\pi t)^{3/2}} {d \over \sinh d}
\end{equation}
where $d=d(x,x')$ is the geodesic distance between $x$ and $x'$.
{\blue
The determinant of the operator $\Delta_0$ on ${\mathbb H_3}$ is then
\begin{align}\label{scalarheatH3}
-\log \left({\det}' \Delta_0\right) 
&= \int_0^\infty {\d t \over t} \int_{{\mathbb H}_3} \d^3x \sqrt{g} \left(K^{\mathbb H_3}_t(x,x) \right)
\cr
&=  {\rm Vol}\left({\mathbb H}_3\right) \int_{0}^\infty {\d t \over t} {e^{-t}\over (4\pi t)^{3/2}}
\cr
&= {1\over 6\pi} {\rm Vol}\left({\mathbb H}_3\right) ~.
\end{align}
In writing the second line we have set $d(x,x)=0$ and pulled out an overall factor of ${\rm Vol} \left({\mathbb H}_3\right)$.  The resulting integral over $t$ diverges, reflecting the usual one-loop ultraviolet divergence.  In the final line we have regulated this divergence by defining the $t$ integral by analytic continuation, as a Gamma function with negative argument.
The final result is proportional to ${\rm Vol} \left({\mathbb H_3}\right)$, and can be interpreted as a one-loop contribution to the bulk cosmological constant.

To apply this to quotients of the form ${\mathbb H}_3/\G$, we first use the method of images to determine the heat kernel
\begin{equation}\label{thesum}
K^{\mathbb H_3/\G}_t(x,x') = \sum_{\gamma \in \G} K^{\mathbb H_3}_t(x,\gamma x')~.
\end{equation}
The determinant of the operator $\Delta_0$ on ${\mathbb H}_3/\G$ is then found by integrating: 
\begin{align}\label{scalarheat}
-\log \left({\det}^\prime \Delta_0\right) 
&= \int{\d t \over t} \int_{{\mathbb H}_3/\G} \d^3x \sqrt{g} \left(K^{\mathbb H_3/\G}_t(x,x) \right)
\cr
&= {1\over 6\pi} {\rm Vol}\left({\mathbb H}_3/\G \right) + \sum_{\gamma \in \G\atop \gamma\ne 1} \int{\d t \over t} \int_{{\mathbb H}_3/\G} \d^3x \sqrt{g} K^{\mathbb H_3}_t(x,\gamma x)~.
\end{align}
In writing the second line we have separated out the $\gamma=1$ term in the sum and computed the $t$ integral as before, again finding a one-loop contribution to the cosmological constant.
To interpret the other term, note that each $\gamma$ can be thought of as an element of the fundamental group, and the heat kernel is just a simple function of the length $d(x,\gamma x)$ of the corresponding geodesic.  
} 
The result is that equation (\ref{scalarheat}) takes the general form of a trace formula -- such as the Selberg or Gutzwiller trace formula -- where the spectrum of a Hamiltonian is related to the lengths of classical orbits. In the present case, the spectrum of $\nabla^2$ is related to the lengths of bulk geodesics.  This will be a general feature of all of our formulas.

The sum over $\G$ is typically impossible to carry out exactly.  It can, however, be simplified by separating out the sum over {\it primitive} elements of $\G$: an element $\gamma\in\G$ is primitive if it cannot be written as a positive  power of any other element of $\G$. 
Each primitive element $\gamma$ generates a ${\mathbb Z}$ subgroup of $\G$ which we call a primitive subgroup (a primitive subgroup has two generators, namely
$\gamma$ and $\gamma^{-1}$).
The sum over $\G$ reduces to a sum over the set ${\cal P}$ of primitive subgroups along with a sum over ${\mathbb Z}$ for each such subgroup.  We note that, since the quotient ${\mathbb H}_3/{\mathbb Z}$ is a solid torus, we can associate to each primitive element $\gamma$ the modular parameter $\tau_\gamma$ of the associated boundary torus.  Writing  $\gamma\in \G\subset \PSL(2,{\mathbb C})$ as a $2\times 2$ matrix, the modular parameter satisfies
% equation for this $\tau_\gamma$ is 
\begin{equation}\label{polo}
2 \cos \pi \tau_\gamma = \Tr\,\gamma.
\end{equation}
This formula does not determine the sign of $\tau_\gamma$, which we fix so that $\mathrm{Im}\,\tau_\gamma>0$.  
It fixes $\mathrm{Re}\,\tau_\gamma$ mod 1 (mod 1 and not mod 2, because lifting from $\PSL(2,\C)$ to $\mathrm{SL}(2,\C)$ means that
 the sign of the right hand side is ill-defined).    But our
subsequent formulas will be expressed in terms of $q_\gamma=e^{2\pi i \tau_\gamma}$, which depends on $\mathrm{Re}\,\tau_\gamma$ only mod 1.

It is now possible to evaluate the integrals $\d t$ and $\d^3x$ in equation (\ref{scalarheat}), and write the result 
as a sum over the set ${\cal P}$ of primitive subgroups.  The result is \footnote{Intermediate steps in this derivation can be found in  \cite{GMY}.}
\begin{equation}
{\det}^\prime\,\Delta_0 = \exp\left\{-{{\rm Vol} \left({\mathbb H_3}/ \G\right)\over 6\pi} \right\} \prod_{\gamma \in {\cal P}} \left(\prod_{\ell,\ell'=0}^\infty \left(1-q_\gamma^{\ell+1} {\bar q}_\gamma^{\ell'+1}\right)\right)^2.
\end{equation}
This is a sum over primitive subgroups, so we do not count $\gamma$ and $\gamma^{-1}$ separately.

Although we have only written the formula for a massless scalar, this procedure can be applied (with more work) to find analogous 
heat kernel expressions for higher spin fields.  The primary difficulty is dealing with the various different tensor structures that 
appear in the heat kernel. We refer to \cite{GMY} for detailed computations.
We will need only the result for the determinant for the spin one Laplacian $\Delta_\mu{}^\nu = - \delta_\mu{}^\nu\nabla^2+ R_\mu{}^\nu$. 
In \cite{GMY}, it was found that\footnote{This formula is not written in precisely this way in \cite{GMY}.    In that reference, $\det\,\Delta_1/\det'\,\Delta_0$
is formally called $\det'\,\Delta_\perp$, and eqn. (\ref{spinone}) is written as a formula for $\det'\,\Delta_\perp$.}
\begin{equation}\label{spinone} \det\,\Delta_1=
{\det}'\,\Delta_0 \cdot \prod_{\gamma \in {\cal P}} \left(\prod_{\ell,\ell'=0}^\infty 
\left(1-q_\gamma^\ell {\bar q}_\gamma^{\ell'+1}\right)\left(1-q_\gamma^{\ell+1} {\bar q}_\gamma^{\ell'}\right)\right)^2.
\end{equation}
(It appears that this formula is valid even if the operator $\Delta_1$ has zero-modes.   In that case, the determinant on the left hand side
of eqn. (\ref{spinone}) vanishes, and the infinite product on the right hand side also vanishes.   This will be discussed in section \ref{discon}.)

We can now assemble these results together to evaluate the expression (\ref{nform}), which formally is the  one-loop determinant of Chern-Simons theory, 
expanded around the trivial flat connection:
\begin{equation}\label{CSdet}
\frac{\left(\det'\,{\Delta_0}\right)^{3/2} }{ {\left(\det\,\Delta_1\right)^{1/2}} }= 
\exp\left\{-{{\rm Vol} \left({\mathbb H_3}/ \G\right)\over 6\pi} \right\} \prod_{\gamma \in {\cal P}} \left(\prod_{n=1}^\infty {1\over |1-q_\gamma^n|}\right) ^2.
\end{equation}

Let us first consider this formula for the solid torus ${\mathbb H}_3/{\mathbb Z}$, where $\G$ is generated by a single primitive element 
$\gamma$ with ${\rm Tr} \left(\gamma\right) = 2 \cos \pi \tau$.  In this particular case, there is only one term in the product on the right hand side of eqn.
(\ref{CSdet}), so this is the only case in which we can evaluate the product in a completely explicit way.  
{\blue
Although the volume of ${\mathbb H}_3 / {\mathbb Z}$ is divergent, we may regularize it using the standard procedures of holographic renormalization.  One begins by
cutting off the volume integral near the boundary and introducing boundary counterterms which remove the divergence that appears as this cutoff is taken away.  The result is not invariant under conformal transformations on the boundary, so requires a choice of metric on the boundary.   With the usual flat metric on the torus, one finds
\begin{equation}
{\rm Vol} \left({\mathbb H}_3 / {\mathbb Z} \right) = - \pi^2\, {\rm Im}\,\tau\,.
\end{equation}
The final result for our the one-loop determinant is\footnote{Similar Chern-Simons computations have appeared in the literature before \cite{Porrati:2019knx}.}
\begin{equation}\label{prodbulk}
\frac{\left(\det'\,{\Delta_0}\right)^{3/2} }{ {\left(\det\,\Delta_1\right)^{1/2}} }= |q|^{-{1/12}}\prod_{n=1}^\infty {1\over |1-q^n|^2} = {1\over |\eta(\tau)|^{2}}~.
\end{equation}
We note that this one loop determinant naturally gives the usual prefactor of $|q|^{-1/12}$, which in the boundary language is attributable to the negative Casimir energy of a free boson on a circle.}
 In bulk gravity calculations, such a term would typically arise from the regularized Einstein-Hilbert action of ${\mathbb H}_3/{\mathbb Z}$.  
In our Chern-Simons computation, this term came for free from the one-loop contribution to the bulk cosmological constant.
Equation (\ref{prodbulk}) is appropriate for $\D=1$; for general $\D$, one simply raises both sides to the power $\D$.
This is the result for the bulk path integral that we needed (eqn. (\ref{judg})) for a bulk theory dual to the average over Narain moduli space.

For higher genus, it is not possible to write such explicit formulas.   But remarkably, it is possible to show in general that if $\Sigma$ is a connected
Riemann surface of  genus $g$, and $Y$ is a handlebody with conformal boundary $\Sigma$, then making the same assumptions, the bulk path
integral on $Y$ agrees with what is needed in eqn. (\ref{poggo}) for the average over Narain moduli space to be reproduced by a sum over handlebodies.

If $Y$ is a genus $g$ handlebody, then its fundamental group $\G$ is a free group on $g$ generators.  Such a subgroup of $\PSL(2,\C)$, acting on ${\mathbb H}_3$
in such a way that the quotient is a handlebody $Y$, is called a Schottky group.    
{\blue
In this case, Zograf \cite{Zograf}, with further developments by
 McIntyre and Takhtajan \cite{McIntyre:2004xs},  proved the following ``holomorphic factorization formula"   %{\blue 
%\begin{equation}\label{tugo}
%\frac{\det {\rm Im}\, \Omega}     {\det'\, \h\Delta_0  } = e^{-S_L} \prod_{\gamma \in {\cal P}} \left(\prod_{n=1}^\infty {1\over |1-q_\gamma^n|^2}\right) 
%\end{equation}
\begin{equation}\label{tugo}
\left(    \frac{\det'\, \h\Delta_0}{\det {\rm Im}\, \Omega}\right)^{-1/2} = e^{S_L/24\pi}%\exp\left\{\frac{S_L}{24 \pi} \right\} 
\prod_{\gamma \in {\cal P}} \left(\prod_{n=1}^\infty {1\over |1-q_\gamma^n|^2}\right) 
\end{equation}
where $\h\Delta_0$ is the Laplacian of a non-compact scalar on the surface $\Sigma$.\footnote{The two dimensional Laplacian $\h\Delta_0$ should not be confused with the three dimensional Laplacian $\Delta_0$ which appeared earlier;  in this section we will denote two-dimensional operators with a hat to avoid confusion.  We note that both $\det'\,\h\Delta_0(\Omega)$ and $S_L$ are not conformally invariant, but rather transform with conformal anomalies
that match in such a way that eqn. (\ref{tugo}) is conformally invariant.  {\blue We also note that we have written (\ref{tugo}) as a product over primitive subgroups ${\cal P}$ where $\gamma$ and $\gamma^{-1}$ are not counted separately.  In the literature this formula is often written in a slightly different way as a product over distinct primitive conjugacy classes, so that $\gamma$ and $\gamma^{-1}$ are counted separately.}}    %}
The function $S_L$ is an appropriately defined Liouville action on the moduli space of Schottky groups, defined explicitly in \cite{ZT}, which plays the role of the 
$|q|^{-1/12}$ factor in the torus case.  Indeed, this Liouville action was proven by Kraznov \cite{Krasnov:2000zq}
to be proportional to the regularized volume of ${\mathbb H}_3/\G$
\begin{equation}\label{regvol}
S_L = -4\, {\rm Vol}\left( {\mathbb H_3}\over \G \right) + {\rm counterterms}~.
\end{equation}
With this result, we see that eqn. (\ref{tugo}) matches exactly the path integral of $U(1) \times U(1)$ Chern-Simons theory on a genus $g$ handlebody given in eqn. (\ref{CSdet}). 
 This includes the factor of $S_L$ that, in other contexts, would arise from a regularized Einstein-Hilbert action.
Recalling the purely two-dimensional version of holomorphic factorization, $\det'\,\h\Delta_0=(\det\,\Im\,\Omega) |\det_{\Gamma_0}'\,\bar\partial|^2$ (see for example
\cite{M,N}; here $\Gamma_0$ is the Lagrangian sublattice associated to the handlebody),
we conclude that our bulk path integral reproduces precisely the desired factor of $|\det_{\Gamma_0}'\,\bar\partial|^{-D}$.

\subsection{Disconnected Boundaries}\label{discon}

Remarkably, rather similar relationships  between bulk and boundary functional determinants continue to hold even when the boundary is disconnected.  
The case that is well-established in the literature is the case that $Y$ is topologically $\Sigma_0\times I$, where $\Sigma_0$ is a surface of genus $g>1$ and $I$
is an open interval.   In this case, if $Y$ is geometrically a quotient $\HH_3/\G$, then the conformal infinity of $Y$ consists topologically of two copies of $\Sigma_0$.
Generically, these two copies, which we will call $\Sigma'$ and $\Sigma''$, have different complex structures.   In that case, the group $\G$ is called
quasi-Fuchsian.   In the special case that the two complex structures are the same, $\G$ is called Fuchsian.    Since the Fuchsian case is just a special
case, we need not consider it separately.

{\blue For quasi-Fuchsian groups,  McIntyre and Teo \cite{McIntyreTeo} showed that for $n>1$,
\begin{equation}\label{McIntyreTeo}
%\left(\frac{ {{\rm Im}~ \det \h\Omega'_n} }{{\det{}' \h\Delta_n(\Sigma') }}\right)
\left(\frac{{\det{}' \h\Delta_n(\Sigma') }}{ {{\rm Im}~ \det \h\Omega'_n} }\right)^{-1/2}
\left(\frac{{\det{}' \h\Delta_n(\Sigma'') }}{ {{\rm Im}~ \det \h\Omega''_n} }\right)^{-1/2}
%\left(  \frac{ {{\rm Im}~ \det \h\Omega''_n} }{{\det{}' \h\Delta_n(\Sigma'') }}\right) 
= e^{{6 n(n-1)+1\over 24\pi}S_L} \prod_{\gamma \in {\cal P}} \left(\prod_{m=n}^\infty {1\over |1-q_\gamma^m|^2}\right) ~.
\end{equation}
}Here $\h\Delta_n$ is the boundary Laplacian acting on a field of spin $n$, and $\Omega_n$ is a generalized period matrix, which reduces to the usual
period matrix if $n=1$ or $n=0$.\footnote{A crucial feature of this formula -- analogous to our use of the bulk geometry in section \ref{disconnected} to determine
an indecomposable Lagrangian sublattice -- is that one is not  free to choose independently the bases of holomorphic cycles on the boundary surfaces in which
 $\Omega'$ and $\Omega''$ are computed.  The choice of bases must be related in a particular way which depends on the bulk geometry \cite{McIntyreTeo}.  In this way the left hand side of this formula depends implicitly on the choice of bulk geometry, consistent with the fact that the right hand side depends on it.}
The left hand side is now interpreted as a product of one-loop determinants  for the two individual boundary theories, regarded as functions of their period matrices.  {\blue
The Liouville action $S_L$ appearing in this equation is again proportional to the regularized volume of ${\mathbb H}_3/\G$, just as in eqn. (\ref{regvol}); this result was established for quasi-Fuchsian groups by Takhtajan and Teo \cite{Takhtajan:2002cc}.
}
%The result of these considerations is that the fluctuations of the $U(1)^{2D}$ Chern-Simons fields in the bulk reproduces the partition function of $D$ free bosons on the boundary, at least in the Schottky and quasi-Fuchsian cases where the relevant holomorphic factorization formulas have been proven.  It is natural to guess that analogous factorization formulas hold for all freely acting discrete groups $\G$, including those with many disconnected boundaries, but we do not know of such results in the literature.

If we could set $n=0$ in this formula, we would be in the same situation as in section \ref{handleb}.   After again invoking eqn. (\ref{CSdet}), we would conclude
 that the product of determinants
on $Y$ agrees with what is needed to reproduce the correlation functions that we studied in section \ref{disconnected} between disconnected boundary
components   (at least for the special case that the two boundaries have the same genus and the indecomposable Lagrangian sublattice considered
is associated to a three-manifold that is topologically $\Sigma_0\times I$).  

In fact, the formula (\ref{McIntyreTeo}) does not hold for $n=0$.   The infinite product on the right hand side is divergent in that case.   We will explain
this in a moment, but first we will point out that this should not be a surprise from the point of view of the bulk Chern-Simons theory.   For the case that the conformal
boundary of $Y$ is the disjoint union of two components $\Sigma'$ and $\Sigma''$, the bulk operator $\Delta_1$ has a zero-mode.   This zero-mode is pure gauge
but it cannot be gauged away by a gauge transformation that is trivial at the boundaries of $Y$.  Being pure gauge, this mode does not contribute to any local
gauge-invariant observable, but it contributes to a Wilson line that stretches between the two conformal boundaries.   To demonstrate the existence of this
zero-mode, one can proceed as follows.   Consider a function $f$ that equals 0 on $\Sigma'$ and equals 1 on $\Sigma''$.   If such a function approaches its
limiting boundary values sufficiently quickly, then 
\be\label{nuff}I(f)=\int_Y\d^3x \sqrt g|\nabla f|^2 \ee
is finite.   By minimizing $I$ within the given class of functions $f$, one can ensure that $\Delta_0 f=0$.   Then $A=\d f$ is a zero-mode of $\Delta_1$
and is square-integrable since $I(f)<\infty$.    Because the operator $\Delta_1$ has this zero-mode, the left hand of eqn. (\ref{spinone}) vanishes, so
we are led to expect that the right hand side will also vanish.  Equivalently, the left and right hand sides of eqn. (\ref{CSdet}) should both be divergent.

The divergence in the infinite product in eqn. (\ref{McIntyreTeo}) -- or equivalently eqn.  (\ref{CSdet}) -- is easiest to see in the Fuchsian case, so
we concentrate on that case.    
The Fuchsian case is the case that  $\G$ sits inside a $\PSL(2,{\mathbb R})$ subgroup of $\PSL(2,{\mathbb C})$.  
In that case, $Y=\Sigma\times I$ has a symmetry that exchanges the two ends of $I$.   The fixed point set of this symmetry is a totally geodesic
embedded surface $\Sigma\cong \HH_2/\G\subset Y$, and all closed geodesics in $Y$ actually lie in $\Sigma$.
So the product in (\ref{McIntyreTeo}) reduces to a product over geodesics in $\Sigma$.  
The Selberg zeta function associated to $\G$ is defined as:
\begin{align}
Z_\G(s) &\equiv \prod_{\gamma \in \P} \prod_{m=0}^\infty \left(1-q_\gamma^{m+s}\right) 
= \prod_{\gamma \in \P} \prod_{m=0}^\infty \left(1-e^{-(m+s) L(\gamma)}\right) 
\end{align}
where $L(\gamma)$ is the length of the geodesic associated to $\gamma$.
We see that the product in (\ref{McIntyreTeo}) is $|Z_\G(1)|^{-2}$.   But it is a standard result that $Z_\G(1)=0$.
This vanishing is equivalent to a divergence $\log Z_\G(s)\to -\infty$ for $s\to 1$. 
The important contributions to $Z_\G(s)$ for $s$ near 1 come from $m=0$ and from very long primitive closed geodesics:
\begin{equation}
\log Z_\G(s)\approx- \int \rho(L) e^{-s L}dL
\end{equation}
where $\rho(L)$ is the density of primitive closed
geodesics with length $L$.   For a compact Riemann surface, we have $\rho(L)\sim {e^L \over L}$ at large $L$,  leading
to a logarithmic divergence in $\log Z_\G(1)$ and vanishing of $Z_\G(1)$.     (We are reversing the usual logic here: the usual procedure is to prove
first in a more direct way thst $Z_\G(1)=0$ and then use this to constrain the large $L$ behavior of $\rho(L$).)

\subsection{$\U(1)^{2\D}$ and $\R^{2\D}$ Chern-Simons Theories}\label{comparison}

The divergence that we have just encountered actually has a simple fix if we take seriously the idea that the bulk theory is a Chern-Simons theory
of the gauge group $\U(1)^{2\D}$.  The Wilson line that stretches between the two boundary components of $Y$ is really valued in the gauge group.
Instead of getting an infinity from a zero-mode of $\Delta_1$, we should get a factor of the volume of the gauge group, which is finite for gauge group
$\U(1)^{2\D}$.   This tells us, then, that we should aim to replace eqn. (\ref{spinone}) and subsequent formulas with a formula in which the zero-mode of $\Delta_1$
 is removed
on the left hand side.  To compensate for this, the heat kernel formulas will have to be modified, and the right hand side of eqn. (\ref{spinone}) would be replaced
with  a regularized version.   Hopefully, there would then also be a regularized version of the McIntyre-Teo formula for $n=0$.   

This particular argument will clearly not work if we assume that the bulk gauge group is $\R^{2\D}$ rather than $\U(1)^{2\D}$.   In this case, the volume of the
gauge group is infinite, and the zero-mode of $\Delta_1$ will really lead to a divergence.

Nonetheless, it seems to be problematical to take too seriously the idea that the bulk theory is a $\U(1)^{2\D}$ gauge theory.  One reason, which does
not depend on the assumed gauge group, was explained
in the introduction:   gauge theory does not tell us to sum over manifolds, and it certainly does not tell us to sum over a specific class of manifolds, such as
handlebodies.   But there is actually a more specific problem if we assume that the theory is a  $\U(1)^{2\D}$ Chern-Simons theory based on
 an even integer unimodular lattice
$\Lambda$.

The $\U(1)^{2\D}$ Chern-Simons theory based on such a lattice is actually completely trivial, in a very strong sense.
As already explained, this  theory is equivalent to $\D$ copies of a $\U(1)^2$ theory with action
\be\label{poj} \frac{1}{2\pi}\int_Y A\d B. \ee
Triviality of this theory is a special case of a statement in \cite{MMS} and was analyzed in considerable detail in \cite{WittenSL}.
Triviality means, first of all, that if $Y$ is an oriented  three-manifold without boundary, then the partition function of the theory on $Y$ is equal to 1.
Second, if $\Sigma$ is a Riemann surface, then the Hilbert space $\H_\Sigma$  of the theory on $\Sigma$ is 1-dimensional, 
and contains a distinguished unit vector $\Psi$.
Third, if $Y$ is any oriented three-manifold with boundary $\Sigma$, then the path integral on $Y$ produces the same vector $\Psi\in\H_\Sigma$.
All of these statements immediately carry over to the $\U(1)^{2\D}$ theory based on an even integer unimodular lattice.

In the case that $Y$ is noncompact, with conformal boundary $\Sigma$, it does not follow from this that the $\U(1)^{2\D}$ Chern-Simons path integral on $Y$
is equal to 1; this depends on what behavior of the fields is assumed near the conformal boundary.    What does follow, however, is that the Chern-Simons path integral 
on $Y$ depends only on $\Sigma$ and not on $Y$.  This may be proved as follows.   Let $Y_0$ be a cutoff version of $Y$ in which the boundary $\Sigma$
is placed at a finite distance rather than at infinity.   And let $U$ be the product of $\Sigma$ with a semi-open interval $[0,1)$.   Thus we can build $Y$ by
gluing together $Y_0$ and $U$ along $\Sigma$.  The dependence of the Chern-Simons path integral on $Y$ is entirely encoded in the vector in $\H_\Sigma$
that is produced by the path integral on $Y_0$.   But this is the same vector $\Psi$, independent of $Y_0$.    Thus no matter what assumption we make
about the behavior of  fields near the conformal boundary, the $\U(1)^{2\D}$ Chern-Simons path integral depends only on the boundary and not on the bulk 
geometry.

What is happening is that $\U(1)^{2\D}$ Chern-Simons places too strong an equivalence relation on manifolds.   In this Chern-Simons theory, all manifolds with given
boundary $\Sigma$  are equivalent.    What we would like instead would be for all manifolds with boundary $\Sigma$ that induce the same Lagrangian sublattice of $\Gamma=H_1(\Sigma,\Z)$
to be equivalent.   It is interesting that we get something very close to this if we just replace $\U(1)^{2\D}$ by $\R^{2\D}$. It suffices
here to consider the basic case $\D=1$ with the two gauge fields $A,B$.    The phase space of $\R^{2}$
Chern-Simons on $\Sigma$ is the tensor product $V= \R^{2}\otimes H^1(\Sigma,\R)$, with a symplectic form that is the tensor product of the quadratic
form $\Lambda$ on $\R^{2}$ and the intersection form on $H^1(\Sigma,\R)$.     An element of $V$ is just a pair $A_0,B_0$ of gauge fields on $\Sigma$
satisfying $\d A_0=\d B_0=0$, up to gauge equivalence.  
The natural gauge-invariant observables in this theory are of the form $\oint_\gamma A$, $\oint_\gamma B$, where $\gamma$ is
a homotopically nontrivial closed loop in $\Sigma$. (As the gauge group is $\R^2$ rather than $\U(1)^2$, these expressions
are gauge-invariant, with no need to exponentiate them.)    Suppose that $\Sigma=\partial Y$ and let $\Psi\in \H_\Sigma$ be the vector
produced by the path integral on $Y$.   Let $\Gamma_0$ be the  Lagrangian sublattice of $\Gamma$ corresponding to $Y$.  
We claim that
for any loop $\gamma\in \Sigma$ whose homology class $[\gamma]$ is in $\Gamma_0$,
\be\label{tofo}\oint_\gamma A\cdot\Psi =\oint_\gamma B\cdot\Psi =0. \ee
Indeed, the condition $[\gamma]\in \Gamma_0$ means that $\gamma$ is the boundary of some oriented
two-manifold
$C\subset Y$, whence $\oint_\gamma A\cdot\Psi=\int_C\d A \cdot \Psi=0$, since $\d A=0$ by the equations of motion.
Similarly $\oint_\gamma B\cdot \Psi=0$.     The operators $\oint_\gamma A$ and $\oint_\gamma B$, for $[\gamma]$ in a Lagrangian
sublatttice $\Gamma_0$,  are a maximal set of commuting observables in this theory, and a state that they annihilate is uniquely
determined by $\Gamma_0$, up to a constant multiple.
Thus any two $Y$'s that induce the same Lagrangian sublattice $\Gamma_0$ generate the same state $\Psi$, up to an overall
constant.  (These
overall constants are not all equal to 1, and for some $Y$'s, the constants in question are divergent because of the infinite volume of the assumed gauge
group.)

We would have preferred to learn that any two $Y$'s associated to the same Lagrangian sublattice determine precisely the same state.  This might
have been an approximation to a statement that in the theory that we are looking for, ``manifolds with boundary $\Sigma$'' are entirely classified by
Lagrangian submanifolds.   However, the $\R^{2\D}$ Chern-Simons theory has come pretty close.

Hopefully by now it is apparent that each of $\U(1)^{2\D}$ and $\R^{2\D}$ have both virtues and vices as candidate gauge groups for a theory dual to
the average over Narain moduli space.

\noindent{\bf Acknowledgments}  

We  thank A. Venkatesh for explanations of matters related
to the Siegel-Weil formula.  We also thank S. Kachru, L. Takhtadjan, Yingkun 
Li, D. Marolf, H. Maxfield, A. McIntyre, G. W. Moore, L. P. Teo, and A. Wall for helpful comments.
Research of AM is supported in part by the Simons Foundation Grant No. 385602 and the
Natural Sciences and Engineering Research Council of Canada (NSERC), funding reference number
SAPIN/00032-2015.  Research of EW is supported in part by  NSF Grant PHY-1911298.  

\appendix

\section{Derivation of the Siegel-Weil Formula at $\D>1$ and $g>1$}\label{diffeq}

In this appendix we will describe in more detail the derivation of the Siegel-Weil formula.   
We will begin in section \ref{narain} by reviewing the moduli space $\M_\D$ of CFTs with $\D$ compact free bosons.  We will show that the torus partition function obeys the differential equation (\ref{zugu}) which was needed in our derivation of the genus one version of the Siegel-Weil formula.
To discuss the higher genus version of this formula, we will first need to review in section \ref{siegel} some facts about the geometry of Siegel upper half space.  We will then discuss the derivation of the higher genus version of the Siegel-Weil formula in section \ref{higherD}.

\subsection{Narain Moduli Space}\label{narain}

A sigma-model with $T^\D$ target space can be described by angle-valued fields $X^p$, $p=1,\dots,\D$ ($X^p\cong X^p+2\pi$), with a metric $G_{pq}$ and
two-form field $B_{pq}$. In this Narain family of conformal field theories, $G$ and $B$ are constants that represent the moduli of the theory.   We will call this moduli space $\M_\D$, and denote a point in $\M_\D$ by $m$.

The action, on a Euclidean signature worldsheet with coordinates $\sigma^\alpha$, $\alpha=1,2$,
flat metric $\delta_{\alpha\beta}$, and Levi-Civita tensor $\varepsilon^{\alpha\beta}$, is
\be\label{pliko} I=\frac{1}{4\pi\alpha'} \int \d^2\sigma\left( G_{pq} \delta^{\alpha\beta}\partial_\alpha X^p\partial_\beta X^q+\i B_{pq}\varepsilon^{\alpha\beta}
\partial_\alpha X^p\partial_\beta X^q\right). \ee
The marginal operator that describes small perturbations $\delta G_{pq}$, $\delta B_{pq}$ of $G$ and $B$ is \be\label{thak} \O=(\delta G_{pq}\delta^{\alpha\beta}+\i \delta B_{pq}
\varepsilon^{\alpha\beta})\partial_\alpha X^p\partial_\beta X^q. \ee  Similarly to the case $\D=1$  discussed in section \ref{practice}, by computing the two-point
function of $\O$ one can determine the Zamolodchikov metric:
\be\label{zamet} \d s^2= G^{mp}G^{nq}\left( \d G_{mn}\d G_{pq}+\d B_{mn}\d B_{pq}\right).   \ee 
This is also the metric of $\M_\D$ as a locally homogeneous space.   
The Laplacian derived from this metric is
\begin{equation}\label{amet}
\Delta_{\M_\D} = 
- G_{mp}G_{nq}\left({\hat \partial}_{G_{mn}}{\hat \partial}_{G_{pq}} + {\hat \partial}_{B_{mn}} {\hat \partial}_{B_{pq}}\right)
-G_{mn}{\hat \partial}_{G_{mn}}
\end{equation}
where ${\hat \partial}_{G_{mn}} = \frac{1}{2}(1+\delta_{mn}){\partial\over\partial{G_{mn}}}$ and ${\hat \partial}_{B_{mn}} = \frac{1}{2}{\partial\over\partial{B_{mn}}}$.

On a torus $\Sigma$ with modular parameter $\tau=\tau_1+\i\tau_2$, the partition function of the model is $Z_\Sigma(m,\tau)=\Theta(m,\tau)/|\eta(\tau)|^2$
where $\Theta(m,\tau)$ is the Siegel-Narain theta function and the denominator does not depend on $m$.   $\Theta(m,\tau)$ is a sum over integer-valued momenta
$n_p$ and windings $w^q$, which we abbreviate as $\vec n$ and $\vec w$.
Explicitly
\be\label{kolo} \Theta(m,\tau)=\sum_{\vec n,\vec w} Q(\vec n,\vec w;m,\tau), \ee
with
\be\label{olo} Q(\vec n,\vec w;m,\tau) =\exp\left( -\frac{\pi \tau_2}{\alpha'}\left(G^{pq} v_pv_q +G_{pq} w^pw^q   \right)  +2\pi\i\tau_1 n_p w^p\right), \ee
where 
\be\label{zolo} v_p=\alpha' n_p+B_{pq} w^q. \ee  
A computation similar to the one in section \ref{practice} but somewhat longer reveals that 
\be\label{reve}\left(\Delta_\H-\D \tau_2 \partial/\partial\tau_2 -\Delta_{\M_\D}\right)Q=0. \ee 
Here $\Delta_\H$, introduced in eqn. (\ref{nugu}), is the Laplacian of the upper half plane.
Since this equation is linear, $\Theta$ satisfies the same equation:
\be\label{poho} \left(\Delta_\H-\D \tau_2 \partial/\partial\tau_2 -\Delta_{\M_\D}\right)\Theta=0.  \ee
The steps that go from this result to the Siegel-Weil formula were explained in section \ref{dtwo}.

\subsection{Geometry of Siegel Upper Half Space}\label{siegel}

We now wish to consider the partition function of the same family of CFTs on a genus $g$ Riemann surface $\Sigma$ with period matrix $\Omega=\Omega_{ij}$, where $i,j=1,\dots, g$.

This period matrix $\Omega$ is an element of Siegel upper half space ${\cal H}_g$, which is the space of complex, symmetric $g \times g$ matrices with positive definite imaginary part:
\begin{equation}
{\cal H}_g \equiv \left\{ \Omega_{ij}: \Omega_{ij} = \Omega_{ji}, {\rm Im}\, \Omega>0\right\}.
\end{equation}
Although not every element of ${\cal H}_g$ can be realized as the period matrix of a Riemann surface, both the Siegel-Narain theta function and relevant Eisenstein series are well-defined functions on ${\cal H}_g$.  This makes the analysis much easier, as ${\cal H}_g$ is considerably simpler than the moduli space of Riemann surfaces.
We will just need to review a few facts about ${\cal H}_g$.

To describe the symplectic group $\Sp(2g,\R)$, we introduce a vector space 
of row vectors
\be\label{colmat}v =\begin{pmatrix}b_1 ~ b_2 \cdots b_g ~ a^1 ~a^2 \cdots a^g\end{pmatrix}, \ee
with matrix elements $ a^1,a^2,\cdots ,a^g$ and $b_1,b_2,\cdots, b_g$
 and a symplectic form $\sum_{i=1}^g\d b_i \d a^i$.  
$\Sp(2g,\R)$ consists of matrices
\be\label{longmat}\gamma=\begin{pmatrix} A& B\cr C & D\end{pmatrix},\ee
constructed from $g\times g $ blocks $A,B,C,D$, that act on $v$ on the right $v\to v\gamma$.   
The condition that $\gamma$ preserves the symplectic form is
\be\label{relns} AB^t=BA^t,~ CD^t =DC^t,~AD^t-BC^t=1. \ee
$\Sp(2g,\R)$ is the group of real-valued matrices that satisfy these conditions.  To get $\Sp(2g,\Z)$, which is known
as the Siegel modular group, we restrict $A,B,C,D$  to be integer-valued.  Likewise we restrict $a^i,b_j$ to be integers, giving an integer lattice 
$\Gamma$ on which $\Sp(2g,\Z)$ acts.   In our application, $\Gamma=H_1(\Sigma,\Z)$.

The  group $\Sp(2g,\R)$ acts on ${\cal H}_g$, by
\begin{equation}
\Omega \to \gamma \Omega \equiv  (A\Omega+B)(C\Omega+D)^{-1}.
\end{equation}
The group $\Sp(2g,{\mathbb Z})$ acts in a proper and discontinuous fashion  on ${\cal H}_g$.  The fundamental domain for this action is complicated (for example at $g=2$ its boundary is a union of 28 pieces) but has finite volume.  The quotient ${\cal A}_g \equiv \Sp(2g,{\mathbb Z}) \backslash {\cal H}_g$ is known as the Siegel modular variety.
When $\Omega$ is the period matrix of a Riemann surface $\Sigma$, this $\Sp(2g,{\mathbb Z})$ action can be thought of as acting on $H_1(\Sigma,{\mathbb Z})={\mathbb Z}^{2g}$.

It is convenient to divide the period matrix into its real and imaginary parts, as $\Omega=x+\i y$.  Since $y_{ij}$ is positive-definite, it is invertible; we will denote its inverse as $y^{ij}$.
The metric on Siegel upper half space
\begin{equation}
\d s^2 = 
y^{ij} y^{kl} (\d y_{ik} \d y_{jl}+\d x_{ik} 
\d x_{jl})
\end{equation}
is Hermitian and invariant under the action of $\Sp(2g,\R)$.
The associated Laplace-Beltrami operator can be written as 
\begin{equation}
\Delta_{{\cal H}_g} = -
y_{ik}y_{jl}({\hat \partial}_{x_{ij}}{\hat \partial}_{x_{kl}} + {\hat \partial}_{y_{ij}} {\hat \partial}_{y_{kl}})
\end{equation}
where ${\hat \partial}_{x_{ij}} = \frac{1}{2}(1+\delta_{ij}){\partial\over\partial{x_{ij}}}$ and ${\hat \partial}_{y_{ij}} = \frac{1}{2}(1+\delta_{ij}){\partial\over\partial {y_{ij}}}$.  This Laplacian commutes with the $\Sp(2g,\R)$ action.

The imaginary part of the period matrix transforms as
\begin{equation}
{\rm Im}\,\Omega= (C{\bar \Omega}+D)^t\,{\rm Im}\,\left(\gamma\Omega\right)\,(C\Omega+D)
\end{equation}
so that 
\begin{equation}
\det {\rm Im}\,\gamma\Omega = {\det {\rm Im}\,\Omega \over  \left| \det (C \Omega+D)\right|^2}\,.
\end{equation}
One can show that
\begin{equation}\label{lapimdet}
\left(\Delta_{{\cal H}_g} + \left(g s^2 - {g(g+1)\over 2} s \right) \right) \left(\det {\rm Im}\,\Omega\right)^s=0.
\end{equation}
The Laplacian commutes with the action of $\Sp(2g,{\mathbb Z})$, so the Eisenstein series
\begin{align}\label{esdef}
E_s(\Omega) 
&\equiv 
\sum_{\gamma\in P\backslash \Sp(2g,{\mathbb Z})}  \left(\det {\rm Im}\,\gamma\Omega\right)^s 
~= ~
\left(\det {\rm Im}\,\Omega\right)^s \sum_{\gamma\in P\backslash \Sp(2g,{\mathbb Z})} \left | \det (C \Omega+D)\right|^{-2s}
\end{align}
is an eigenfunction of the Laplacian with the same eigenvalue.
This is the Eisenstein series which appears in (\ref{wilpo}) as the average CFT partition function on a genus $g$ surface.
In this equation we have defined the Siegel parabolic subgroup $P\equiv \left\{\left({A~B\atop 0~D}\right)\in \Sp(2g,{\mathbb Z})\right\}$; this is the subgroup of $\Sp(2g,{\mathbb Z})$ which acts trivially\footnote{Here one has to know that $\det\,D=\pm 1$, as explained
shortly.}  on $|\det {\rm Im}\,\Omega|$.  It follows that $E_s(\Omega)$ is invariant under the action of $\Sp(2g,{\mathbb Z})$.
We note that the sum diverges when ${\rm Re}\,s\leq\frac{g+1}{2}$.   The Eisenstein series $E_s(\Omega)$ can be analytically continued from the region of convergence to define a meromorphic function in the whole complex $s$-plane, but its direct relation to an
average over Narain moduli space only holds for ${\rm Re}\,s>\frac{g+1}{2}.$

\def\a{{\mathbf a}}
In eqn. (\ref{belfo}), we defined the Eisenstein series in a seemingly different way as 
 a sum over Lagrangian sublattices.   The relation between
the two definitions is as follows.   First of all, the condition $b_i=0$ defines a particular Lagrangian sublattice 
$\Gamma_0\subset
\Gamma$.    The subgroup of $\Sp(2g,\Z)$ that leaves $\Gamma_0$ fixed is precisely the Siegel parabolic group $P$.
So the sum over  $P\backslash \Sp(2g,{\mathbb Z})$ is precisely the sum over Lagrangian sublattices.  Concretely, 
$\Gamma_0$ is spanned by row vectors
$(0,\a )$, where $\a=(a^1,a^2,\cdots, a^g)$ is a $g$-component row vector.  For $\gamma=\begin{pmatrix} A & B\cr C & D\end{pmatrix}\in\Sp(2g,\Z)$,
we have \be\label{wiiffo} (0,\a)\gamma= (\a C, \a D).\ee
   Thus $\gamma$ maps $\Gamma_0$ to a new Lagrangian sublattice spanned by
vectors $(\a C,\a D)$ for arbitrary $\a$.   So the Eisenstein series can be written as a sum over the pairs $(C,D)$, 
subject to the equivalence relation $(C,D)\cong (UC,UD)$, $U\in {\mathrm{GL}}(g,\Z)$, which comes from the action of $P$.

In proving that $P$ is the automorphism group of the Lagrangian sublattice $\Gamma_0$, there is just one
nontrivial point.   It is immediate that an element  $\gamma=\left({A~B\atop 0~D}\right)\in P$ maps $\Gamma_0$ to itself, 
but for it to be an automorphism of $\Gamma_0$ (as an integer lattice), one needs  $\det\,D=\pm 1$.     
In fact, any element of the symplectic group has determinant 1, and for the block triangular matrix $\gamma$ we have
$\det\,\gamma=\det\,A\det \,D$.  Since $\det\,A$ and $\det\,D$ are integers, the fact that $\det\,\gamma=1$ implies
that $\det\,A$ and $\det\,D$ are both $\pm 1$. This condition is equivalent to $A$ and $D$ having integer-valued inverses, and
thus belonging to $\mathrm{GL}(g,\Z)$.
  Actually for $\gamma\in P $, $A$ and $D$ 
can be arbitrary elements of $\mathrm{GL}(g,\Z)$, constrained by $AD^t=1$.  

The group $ P $ also contains matrices  $\begin{pmatrix} 1& B\cr 0 & 1\end{pmatrix}$, where the
only constraint on $B$ is that it is symmetric and integer-valued.
 Such group elements act 
 on the period matrix by $\Omega\to \Omega+B$, thus shifting ${\mathrm{Re}}\,\Omega$ by an arbitrary symmetric
 integer-valued matrix,
 and leaving $\Im\,\Omega$ fixed.

\subsection{The Average CFT Partition Function at Genus $g$}\label{higherD}

We now consider the sigma-model with $T^\D$ target space on a Riemann surface of genus $g$.
The partition function is a function of both the Narain moduli and the moduli of the Riemann surface.  As in the torus case, the partition function is equal to a Siegel-Narain theta function times an oscillator contribution.  The oscillator contribution is independent of $\M_\D$, so will factor out when we average over $\M_\D$. 

The genus $g$ version of the Siegel-Narain theta function depends on both $m\in \M_\D$ and the period matrix $\Omega \in {\cal H}_g$ of our Riemann surface, and can be written as
 \begin{equation}\label{longdef}
 \Theta(m,\Omega)= \sum_{{\bf n}, {\bf w} \in \Z^{g\times p}} Q({\bf n},{\bf w},m,\Omega)
 \end{equation}
where
  \begin{equation}\label{longsum}
Q({\bf n},{\bf w},m,\Omega)\equiv 
\exp\left\{-{\pi y_{ij}\over \alpha'} \left(G^{pq} v^i{}_p v^j{}_q+ G_{pq} w^{ip} w^{jq}\right) +2\pi i x_{ij} n^i{}_p w^{jp}\right\}
 \end{equation}
with 
 \begin{equation}
 v^i{}_p = \alpha' n^i{}_p + B_{pq}w^{iq}.
 \end{equation}
Note that the momentum ${\bf n}=n^i{}_p$ and winding ${\bf w}= w^{ip}$ are now $g\times p$ matrices, with $i=1,\dots, g$ and $p=1,\dots, \D$.
The Siegel-Narain theta function is not modular invariant, but rather transforms under $\Sp(2g,{\mathbb Z})$ transformations in such way that
$\left(\det {\rm Im}\,\Omega\right)^{\D/2}\Theta(m,\Omega)$ is invariant.  It will therefore be convenient to work with this combination of the determinant and the theta function.
 
The starting point for our derivation of the Siegel-Weil formula is the following differential equation for $Q$:
 \begin{equation}\label{masterEq}
 \left(\Delta_{{\cal H}_g} -\Delta_{\M_\D}+ \frac{g \D(\D-g-1)}{4}\right) \left( \left(\det {\rm Im}\,\Omega\right)^{\D/2}Q({\bf n},{\bf w},m,\Omega)\right)=0.
 \end{equation}
The derivation of this differential equation from our previous expressions for the Laplacians is somewhat lengthy.  So we will just make a few comments on its derivation.  The first is that our formulas for $\Delta_{{\cal H}_g}$ and $\Delta_{\M_\D}$ are quite similar to one another.  So it is perhaps not surprising that many of the terms which appear when $\Delta_{{\cal H}_g} -\Delta_{\M_\D}$ acts on $Q$ directly cancel with one another.  There are additional terms which come from  (among other things) the piece of  $\Delta_{\M_\D}$ which is linear in ${\hat \partial}_{G_{pg}}$ acting on $Q$, but these cancel against the terms that are linear in $\partial_{y_{ij}}$ acting on $ \left(\det {\rm Im}\,\Omega\right)^{\D/2}$.  
This just leaves the terms where all of the derivatives in $\Delta_{{\cal H}_g}$ act on 
$ \left(\det {\rm Im}\,\Omega\right)^{\D/2}$, which gives the constant term in (\ref{masterEq}), according to equation (\ref{lapimdet}).

We now perform the sum over ${\bf n}$ and ${\bf w}$ to get
\begin{equation}
 \left(\Delta_{{\cal H}_g} -\Delta_{\M_\D}+ \frac{g \D(\D-g-1)}{4}\right) \left( \left(\det {\rm Im}\,\Omega\right)^{\D/2}\Theta(m,\Omega)\right)=0.
 \end{equation}
We define $\W(\Omega)\equiv \left(\det {\rm Im}\,\Omega\right)^{\D/2} F(\Omega)$, where 
 \begin{equation}
 F(\Omega) =
 \int_{\M_\D} \Theta(m,\Omega)  \,\d\mu(m)
 \end{equation}
 is the average lattice theta function.    
We may then use the fact that, for sufficiently large $\D$, the 
integral $ \int_{\M_\D} \Delta_{\M_\D} \Theta(m,\Omega)\d\mu(m)$ vanishes to conclude that 
 \begin{align} 
 \left(\Delta_{{\cal H}_g} + \frac{g \D(\D-g-1)}{4}\right) \W(\Omega) =0.
  \end{align}
The result is that $\W(\Omega)$ is an $\Sp(2g, {\mathbb Z})$ invariant function which obeys precisely the same eigenvalue equation %on ${\cal H}_g$ 
as $E_{\D/2}(\Omega)$.
% are eigenfunctions of $\Delta_{{\cal H}_g}$ with eigenvalue $\frac{1}{4}g D(D-g-1)$.

We will now take $\D>g+1$, so that the eigenvalue of $\Delta_{{\cal H}_g}$ is negative. 
In this case $ \W(\Omega)$ and $E_{D/2}(\Omega)$ must be equal.    We explained the proof of this step for genus $g=1$ in 
section  \ref{dtwo}: one shows that the difference $E_{\D/2}(\Omega)- \W(\Omega)$ is square-integrable, and therefore as an eigenfunction of the Laplacian with a negative eigenvalue, it must vanish.   The proof for arbitrary $g$
is similar but technically more complicated.   We will only provide a sketch.

What makes the case of general $g$ more complicated is that there are different ways that $\Omega$ can go to infinity.
Of course, the inequivalent  possibilities are somewhat limited by the $\Sp(2g,\Z)$ symmetry.   Because of
the symmetry of shifting $\Omega$ by an arbitrary integer-valued symmetric matrix (see the final comment of section \ref{siegel}),
there is no meaningful notion of ${\mathrm{Re}}\,\Omega$ becoming large, and we can keep it fixed in the following discussion.
Similarly, we do not have to worry about the possibility that an eigenvalue of $y=\Im\,\Omega$ becomes small (thus reaching
the boundary of  the Siegel upper half space $\H_g$);  by an $\Sp(2g,\Z)$ transformation we can map any limiting behavior of $y$
to the possibility that $y$ is becoming large.   However, there are different ways for $y$ to become large and we have to be careful
about this.

Looking back to the lattice sum (\ref{longdef}) that enters the definition of the Siegel-Narain theta function, we see that when
$y$ becomes large, some contributions to the sum over the $g$-plets of integers $n^i{}_p$ and $w^{jq}$ are strongly suppressed.   
For example, if $y$ becomes large in a completely generic way, all of its eigenvalues becoming large, then
all contributions  are strongly suppressed unless ${\bf n}={\bf w}=0$.   At the other
extreme, if $y_{11}$ becomes large while other matrix elements of $y$ remain fixed, then the surviving contributions in the lattice
sum are those with $n^1{}_p=w^{1q}=0$, but no constraint on the other integers in the lattice sum.   In general, there are $g$
essentially different ways for $y$ to go to infinity.   An example of the $k^{th}$ possibility is that the large matrix elements of $y$
might be $y_{11},y_{22}, \cdots , y_{kk}$.   A more general way to describe this situation is to say that the $k^{th}$ possibility
is that $y$ goes to infinity in such a way that the lattice sum in eqn. (\ref{longdef}) is   reduced to a sum over $(g-k)$-plets of integers,
for some $k\in \{1,2,3,\cdots, g\}$.   In terms of Riemann surfaces, what is happening is that a genus $g$ surface is degenerating
to a surface of genus $g-k$, with $k$ pairs of points glued together.

We will first consider the case that $k=g$, which we will describe by saying that $y$ is uniformly large. 
 First let us look at the Eisenstein series (\ref{esdef}).   We see immediately see that for generic large $y$ (with fixed 
$x={\mathrm{Re}}\,\Omega$), the contribution to $E_s(\Omega)$ with $C=0$ is $(\det\,\Im\,\Omega)^s$, while any other contribution is of
order $1/(\det\,\Im\,\Omega)^s$.  (There is only one contribution with $C=0$, because the condition $C=0$ means
that $\gamma\in P $, regardless of $D$.)      So for uniformly large $y$,
\be\label{okj} E_s(\Omega)\sim (\det\,\Im\,\Omega)^s +\O((\det\,\Im\Omega)^{-s}).\ee    Now let us look at the definition of the
Siegel-Narain theta function in eqn. (\ref{longdef}).   With $y$ large and generic, all contributions to the lattice sum with nonzero
integers ${\bf n}$, ${\bf w}$ are strongly suppressed.    Thus the Siegel-Narain theta function reduces to $\Theta(m,\Omega)=1$.
After averaging this over $m$, we get $F(\Omega)=1$, and hence $ \W(\Omega)=(\det\,\Im\Omega)^{\D/2}$, plus corrections
that vanish when $y$ becomes uniformly large.   So we have confirmed that in this region, $E_{D/2}(\Omega)$
and $ \W(\Omega)$ coincide, modulo terms that vanish asymptotically.\footnote{\label{delicate} A detail about these subleading terms
 might be puzzling at first.   In  eqn. (\ref{okj}), we see that the subleading terms in $E_{D/2}(\Omega)$ are power law
suppressed when $y$ becomes uniformly large, 
while from eqns. (\ref{longdef}) and (\ref{longsum}), it may appear that contributions to the theta function
with ${\bf n}$ or ${\bf w}$ nonzero are exponentially suppressed for uniformly large $y$.  
That last statement is true for fixed values of the Narain
moduli $G,B$.   However, we are really interested in averaging over these moduli to get $F(\Omega)$.   In this averaging,
it is possible for $G$ to be very large.    For uniformly 
large $y$, the averaged theta function has contributions that decay like a power of $y$  that come from ${\bf n},\,{\bf w}\not=0$
 but  $G\sim y$.   This gives power law suppressed contributions to the averaged theta function
 because the measure of Narain moduli space
 decays as a power of $G$ for large $G$.}   

\def\b{{\mathbf b}}
Finally we will discuss what happens when $y$ becomes large in a nonuniform fashion.   For illustration, we consider the case $k=1$.
The other cases are similar.  For $k=1$, we may assume that the only large matrix element of $y$ is $y_{11}$.   Looking back
to eqn. (\ref{esdef}), we see that all contributions to $E_s(\Omega)$ are suppressed in this region except those
with $C_{i1}=0$, $i=1,\cdots,g$.   Since one of the columns of $C$ vanishes, it follows that there is a nonzero $g$-component row
vector $\a_0$ with $\a_0 C=0$. We can choose $\a_0$ to be primitive.
  The Lagrangian sublattice $\Gamma_0$ associated to the pair $(C,D)$ then contains the vector $(0,\a_0 D)$.
Using the equivalence relation $(C,D)\cong (UC,UD)$, $U\in {\mathrm{GL}}(g,\Z)$, which does not affect the condition $C_{i1}=0$,
we can put $\a_0 D$ in the form $\a=(1,0,0,\cdots,0)$.  Once we do this,  $\Gamma_0$  contains
the vector $(\b,\a)$, with $\b=(0,0,\cdots,0)$.
    $\Gamma_0$ is spanned by this vector together with a rank $g-1$
lattice $\Gamma_0'$ of vectors $(\b,\a)$ where $\b $ and $\a$ have vanishing first component: $\b=(0,b_2,\cdots, b_g)$ and $\a=(0,a^2,\cdots, a^g)$. 
But $\Gamma_0'$ is just a Lagrangian sublattice of $\Z^{2g-2}$.   Thus for $y_{11}$ large, $E_s(\Omega)$ reduces
to $(\det\,\Im\,\Omega)^s$ times a sum over Lagrangian sublattices of $\Z^{2g-2}$,   plus terms that vanish for $y_{11}\to\infty$.
Now let us compare this to the averaged theta function.    In making this comparison, we assume inductively that
we already know that $E_{D/2}(\Omega)=\W(\Omega)$ for genus less than $g$, and we will prove that for genus $g$,
$E_{D/2}(\Omega)-\W(\Omega)$ vanishes for $y_{11}\to\infty$.   (As explained earlier, if $\D>g+1$, it then follows that $E_{D/2}(\Omega)=
\W(\Omega)$ in genus $g$.)  For this, we just observe
from eqns. (\ref{longdef}) and (\ref{longsum})  that for $y_{11}\to\infty$, the surviving
contributions to the lattice sum that define the theta
function are those with $n^1{}_p=w^{1p}=0$, so that this lattice sum reduces to a sum of the same form with $g$ replaced by
$g-1$.  By the inductive hypothesis, the average of this restricted sum is related to the Eisenstein series with $g$ replaced by $g-1$.\footnote{From the point of view of conformal field theory, the meaning of this is that when a handle of a Riemann surface becomes
very long (something which is equivalent conformally to the degeneration from genus $g$ to genus $g-1$), the dominant
contribution to the partition function
comes from the vacuum state flowing through the long handle.   This contribution is the partition function of the same theory
on a surface of genus $g-1$.}  So $ \W(\Omega)$ for genus $g$
agrees for $y_{11}\to \infty$ with $E_{D/2}(\Omega)$.     As in footnote \ref{delicate}, to compare
terms in $E_{D/2}(\Omega)$ and in $ \W(\Omega)$ that vanish for $y_{11}\to\infty$, one must take into account the behavior at large 
$G$, which gives the dominant correction at large $y_{11}$ on the CFT side.

\bibliographystyle{unsrt}

\end{document}